\documentclass[twocolumn,showpacs,preprintnumbers,amsmath,amssymb,prb,superscriptaddress]{revtex4-1}
\pdfoutput=1
\usepackage{graphicx}
\usepackage{dcolumn}
\usepackage{float}
\usepackage[tight]{subfigure}
\usepackage{amsmath}
\usepackage{verbatim}
\usepackage{units}
\usepackage[usenames,dvipsnames]{color}
\usepackage{bm}
\usepackage[normalem]{ulem}

\usepackage{pdfpages}

\newcommand{\veco}[1]{ \hat{ \mathbf{#1}} }
\newcommand{\bra}[1]{\langle #1|}
\newcommand{\ket}[1]{|#1 \rangle}
\newcommand{\braket}[1]{\langle #1 \rangle}
\newcommand{\suppinfo}{Supplementary Information}

\newcommand{\Eq}[1]{equation~(\ref{#1})}
\newcommand{\eq}[1]{(\ref{#1})}
\newcommand{\Fig}[1]{Fig.~\ref{#1}}

\begin{document}
\title{Spintronic magnetic anisotropy}
\author{Maciej Misiorny}
\affiliation{Peter Gr{\"u}nberg Institut, Forschungszentrum J{\"u}lich, 52425 J{\"u}lich,  Germany}
\affiliation{JARA\,--\,Fundamentals of Future Information Technology}
\affiliation{Faculty of Physics, Adam Mickiewicz University, 61-614 Pozna\'{n}, Poland}
\author{Michael Hell}
\affiliation{Peter Gr{\"u}nberg Institut, Forschungszentrum J{\"u}lich, 52425 J{\"u}lich,  Germany}
\affiliation{JARA\,--\,Fundamentals of Future Information Technology}
\author{Maarten R. Wegewijs}
\affiliation{Peter Gr{\"u}nberg Institut, Forschungszentrum J{\"u}lich, 52425 J{\"u}lich,  Germany}
\affiliation{JARA\,--\,Fundamentals of Future Information Technology}
\affiliation{Institute for Theory of Statistical Physics, RWTH Aachen, 52056 Aachen,  Germany}

\maketitle
\textbf{
 An attractive feature of magnetic adatoms and molecules for nanoscale
 applications is their superparamagnetism, the preferred alignment of
 their spin along an easy axis preventing undesired spin reversal.
 The underlying magnetic anisotropy barrier -- a quadrupolar energy
 splitting -- is internally generated by spin-orbit interaction and can
 nowadays be probed by electronic transport.
 Here we predict that in a much broader class of quantum-dot systems with
 spin larger than one-half, superparamagnetism may arise \emph{without}
 spin-orbit interaction:
 by attaching ferromagnets a \emph{spintronic} exchange field of
 \emph{quadrupolar} nature is generated locally. It can be observed in
 conductance measurements and surprisingly leads to enhanced spin
 filtering even in a state with zero average spin. Analogously to the
 spintronic dipolar exchange field, responsible for a local spin torque,
 the effect is susceptible to electric control and increases with tunnel
 coupling as well as with spin polarization.
}

%%%%%%%%%%%%%%%%%%%%%%%%%%%%%%%%%%%%%%%%%%%%%%%%%%%%%%%%%%%%%%%%%%%%%%%%%%%%%%%%%%%%%%%%%%%%%%%
The growing interest in nanomagnets, e.g., magnetic adatoms~\cite{Brune_Surf.Sci.603/2009} and single-molecule magnets~\cite{Gatteschi_book}
 is fueled by prospects of their application in novel spintronic devices whose functionality derives from their unique magnetic features~\cite{Bogani_NatureMater.7/2008}.
A key property of such systems is their strong magnetic anisotropy
leading to  magnetic bistability, required for building blocks for nanoscale memory cells~\cite{Mannini_NatureMater.8/2009,Loth_NaturePhys.6/2010}
and non-trivial quantum dynamics, useful for  quantum information processing~\cite{Leuenberger_Nature410/2001,Tejada_Nanotechnology12/2001}.
In either case, operational stability of such devices hinges heavily on the height of the energy barrier opposing the spin reversal.
Though recently progress in the control over the magnetic anisotropy by synthesis~\cite{ChemSocRev:MolBasedMag}, mechanical straining~\cite{Parks_Science328/2010}, atomic manipulation~\cite{Otte_NaturePhys.4/2008} or electrical gating~\cite{Zyazin_NanoLett.10/2010} has been made,
 achieving of a high spin-reversal barrier still remains a challenge.
Incorporating a nanomagnet  into an electronic circuit  may significantly alter its magnetic properties~\cite{Mannini_Adv.Mater.21/2009,Rogez_Adv.Mater.21/2009,Kahle_NanoLett.12/2012},
but may also be advantageous. One possible, {spintronic} route for manipulation of nanomagnets entails ferromagnetic electrodes and uses the spin torque due to spin-polarized scattering~\cite{Maekawa_book}  or Coulomb interaction~\cite{Konig_Phys.Rev.Lett.90/2003},
 magnetic analogs of the proximity effect in superconducting junctions.
In this article,  we present another route that combines spintronics with molecular magnetism:
high-spin quantum dots can acquire a significant magnetic anisotropy that is purely of \emph{spintronic} origin, instead of deriving from the spin-orbit interaction, as
the tunneling to ferromagnets induces a local, \emph{quadrupolar exchange field}.
Besides providing an alternative approach to electrical manipulation and engineering of superparamagnetic nanomagnets, this new quantity is of key importance for the analysis of experiments that probe atoms or molecules using highly spin-polarized electrodes.

%%%%%%%%%%%%%%%%%%%%%%%%%%%%%%%%%%%%%%%%%%%%%%%%%%%%%%%%%%%%%%%%%%%%%%%%%%%%%%%%%%%%%%%%%%%%%%%

\begin{figure*}[t]
  \includegraphics{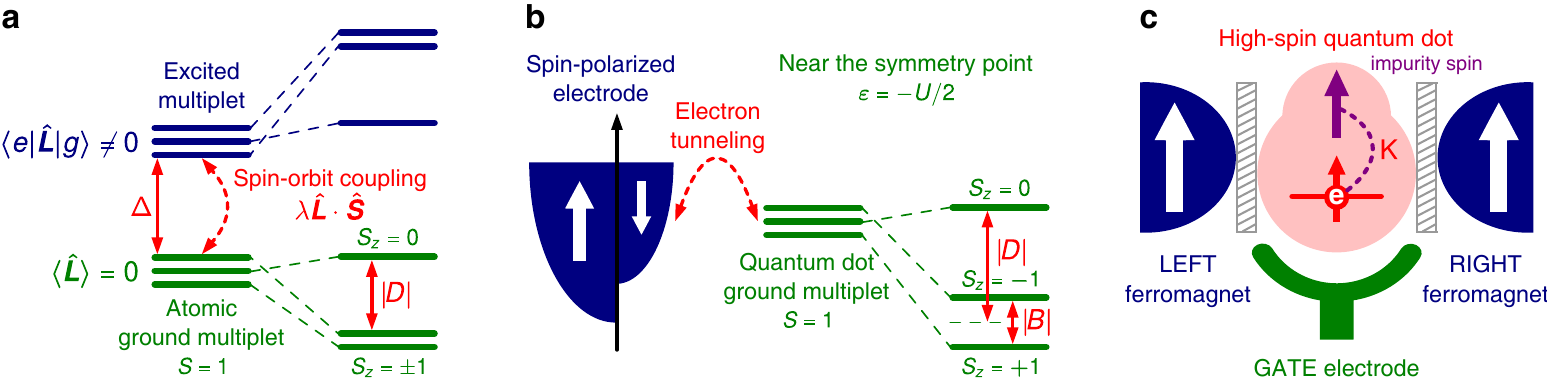}
  \caption{
    \textbf{The origin of magnetic anisotropy splitting of a high-spin ground multiplet}.
    \textbf{a}, The atomic case, a spin-1 multiplet with quenched orbital moment with virtual spin-orbit scattering into an excited state.
    \textbf{b}, The spintronic case, a spin-1 quantum dot with virtual electron tunneling into an attached ferromagnet.
    \textbf{c}, Generic model of a high-spin $S=1$ quantum-dot spin valve (see Methods).
  }
  \label{fig:1}
\end{figure*}

\begin{figure}[t]
   \includegraphics{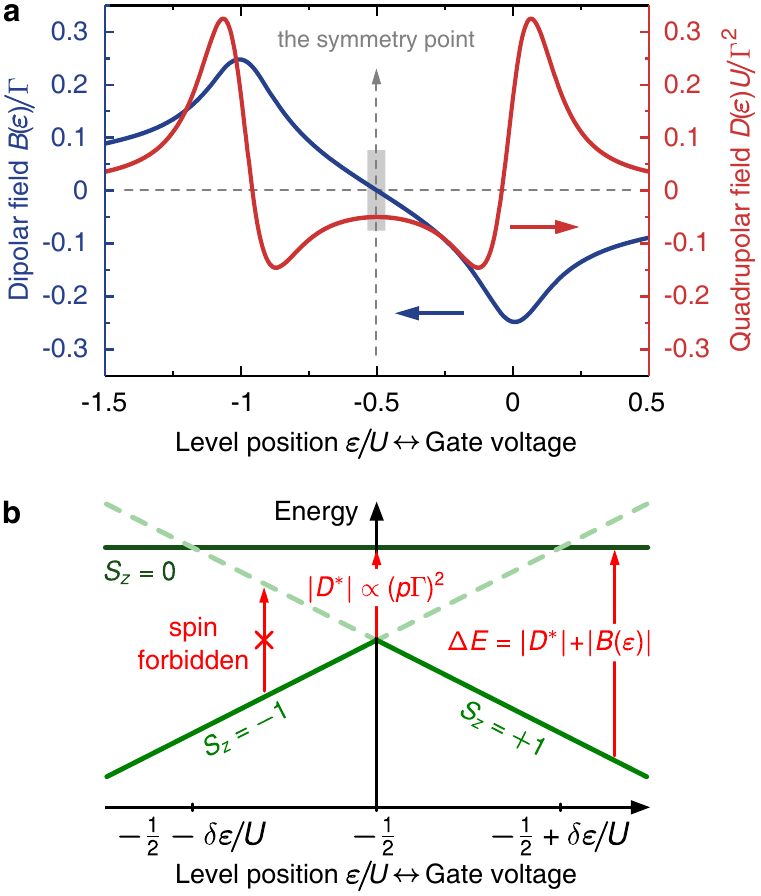}
  \caption{
    \textbf{Effective exchange fields}. (Real-time perturbation theory)
    \textbf{a}, Dipolar $B(\varepsilon)$ and quadrupolar $D(\varepsilon)$ exchange fields
    as a function of the quantum dot level position ($-\varepsilon\propto V_\textrm{g} =$ gate voltage) at zero bias voltage.
     Parameters: $W=1$ eV, $U=100$ meV, $\Gamma/U=0.01$, $T/U=0.05$  and $p=0.5$.
     \textbf{b},~Excitation spectrum of a nanomagnet around the symmetry point as generated by the magnetic proximity effect.
  }
  \label{fig:2}
\end{figure}

\begin{figure*}[t]
  \includegraphics{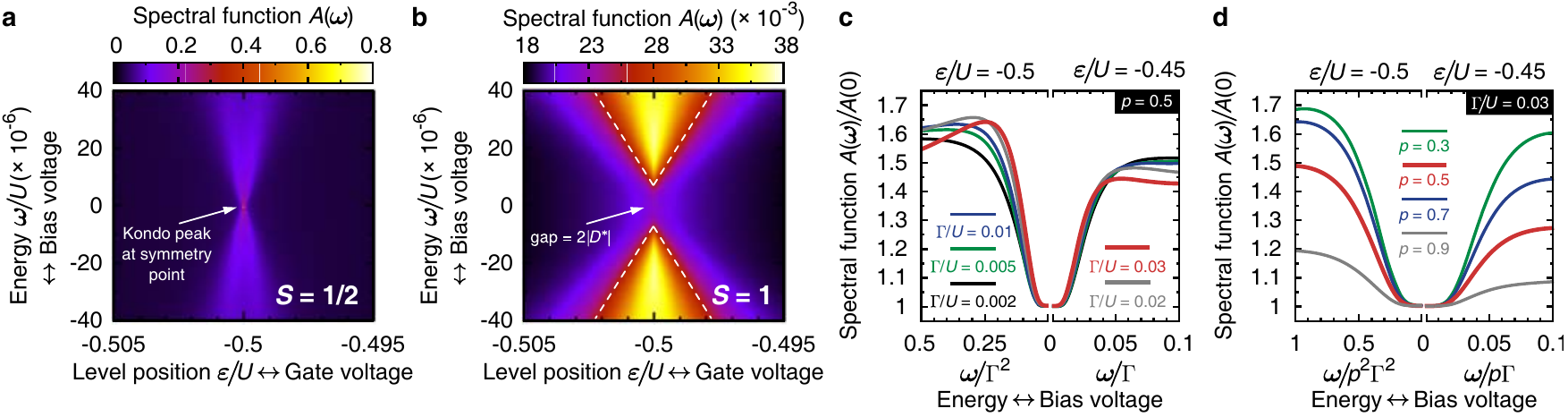}
  \caption{
    \textbf{Spectroscopic features of spintronic anisotropy}. (DM-NRG)
    Dependence of the differential conductance $\textrm{d}I/\textrm{d}V_\textrm{b}$ at $V_\textrm{b}=\omega$, approximated by
    the equilibrium spectral function~$A(\omega)$, on the  energy $\omega$ and
    the level position $\varepsilon$    close to the center of the Coulomb blockade regime,   $\varepsilon=-U/2$.
    \textbf{a}-\textbf{b}, Spectral function for parallel ferromagnets with spin polarization $p=0.5$ coupled to
    a spin $S=1/2$ single-level quantum dot ($K=0$) in \textbf{a}, and
    a spin $S=1$  quantum dot ($K<0$) in \textbf{b}.
    Dashed lines in \textbf{b} represent the gap $\Delta E$ (see \Fig{fig:2}b and supplementary figure S-2).
    \textbf{c}-\textbf{d},
    For the $S=1$ case relevant $\omega$-cross-sections for several different values of
     the tunnel coupling $\Gamma$ (\textbf{c}) and
     the spin polarization $p$ of the ferromagnets (\textbf{d}) are shown.
    In both \textbf{c} and \textbf{d}, the left/right side of the $\omega$-axis corresponds to the regime dominated by
    the quadrupolar field,
    $|B(\varepsilon)| \ll |D(\varepsilon)|$, and
    the dipolar field,
    $|B(\varepsilon)| \gg |D(\varepsilon)|$, respectively.
    Parameters: $W=1$ eV, $U=100$ meV, $K/U=-0.01$,  $T=0$, and in (a)-(b) $\Gamma/U=0.03$.
  }
  \label{fig:3}
\end{figure*}

%================================================================================================
\section*{Dipolar vs. quadrupolar exchange field}

The origin of  superparamagnetism, usually dominating  the magnetic behaviour of a nanomagnet, is a magnetic anisotropy energy barrier.
For instance, an adatom with a spin-degenerate ground multiplet (quenched orbital moment), is described by the generic \mbox{spin~Hamiltonian~\cite{Gatteschi_book}}
    \begin{equation}
    \mathcal{H}_\text{eff} =   B\hat{S}_z + D \hat{Q}_{zz}
    ,
    \label{eq:heff}
    \end{equation}
where $\hat{S}_z$ denotes the component of the total spin ($S\geqslant1$) along the $z$-axis,
and $\hat{Q}_{zz}=\hat{S}_z^2 -\tfrac{1}{3}S(S+1)$ is a diagonal component of the spin-quadrupole moment tensor. Moreover,
 $B$ and $D$ are dipolar and quadrupolar fields, respectively.
All parameters are in units of energy ($k_\textrm{B}=\hbar=|e| = 1$).
If $D<0$, the quadrupolar term prefers the axial spin states over the planar ones, i.e. the spin is aligned with the $z$-axis but without favouring a particular orientation along it.
For spin $S=1$ the corresponding energy splitting is sketched in \Fig{fig:1}a. At temperatures $T<|D|$, it prevents transitions between the axial spin states \emph{via} an intermediate planar state (i.e. spin reversal), while maintaining the former ones as ground states.  This superparamagnetism is thus of major interest for applications in which the axial states represent an information bit, and such transitions are unwanted. On the contrary, the first term in \Eq{eq:heff}, coupling the spin dipole to the \emph{external} magnetic field $B$ (chosen along the $z$-axis), does introduce  a distinction between the `up' and  `down' axial states.
The crucial role of the $z$-axis stems from the second term of~\Eq{eq:heff}. It emerges~\cite{Gatteschi_book}
when taking into account virtual scattering within the ground state multiplet $\ket{g}$ of the adatom through the high-energy excited state $\ket{e}$ at energy $\Delta$,
caused by the spin-orbit interaction
$
\mathcal{H}_\text{s-o} =  \lambda \veco{L} \cdot \veco{S}
,
$
see \Fig{fig:1}a.
Often, only one component of orbital angular momentum has non-zero matrix elements due to  ligand-field hybridization and
an uniaxial \emph{intrinsic} anisotropy along the $z$-axis is imposed by the negative
$
D=-\tfrac{3}{2}\lambda^2
|\bra{g} \hat{L}_z \ket{e}|^2/\Delta
$.
Hence, by probing the ligand environment, the atomic electrons experience a broken spin symmetry.

In spintronics a very similar situation, depicted in \Fig{fig:1}b,  arises in a completely different physical setting where spin-orbit interaction is \emph{negligible}.
Electrons localized in a high-spin ($S>1/2$), spin-isotropic  quantum dot  probe the broken spin symmetry in attached ferromagnets by
virtual charge fluctuations. These fluctuations result in a spin current, which
transfers spin angular momentum from the electrode to the quantum dot,
resulting in a spin torque. It can be described by \emph{replacing} the externally applied magnetic field in~\Eq{eq:heff}
 by an effective \emph{exchange field}.
Specifically, we consider the setup outlined in \Fig{fig:1}c:
a quantum dot with a triplet ground state obtained by coupling an orbital level  \emph{via} Heisenberg interaction $(K < 0)$ to an immobile, spin-one-half impurity,
 causing a singlet-triplet splitting $|K|$.
The orbital level is additionally tunnel-coupled to the ferromagnets.
None of the simplifying assumptions made on the impurity, namely its immobility and its low spin value, is crucial for what follows, see \suppinfo.
The physics can be understood by first ignoring the second ferromagnet.
Then, the \emph{dipolar exchange field} $B$ to the leading order in the tunneling rate $\Gamma$ decomposes into a difference of two contributions $B_0$ and $B_2$~\cite{Sothmann_Phys.Rev.B82/2010}:
    \begin{equation}
    B  = B_0 - B_2
    \ \
    \textrm{and}
    \ \
    B_n  =
    \mathcal{P}\!\!
    \int_{-W}^{W} \!
    \frac{\textrm{d}\omega}{2 \pi}\,\frac{p \Gamma f(\omega)}{\omega-\varepsilon-nU/{2}},
    \label{eq:B}
    \end{equation}
with $\mathcal{P}$ standing for the principal value integral.
Here, $f(\omega) \!=\! \big[\textrm{e}^{\omega/T }\! +\! 1 \big]^{-1}$, and $\varepsilon$ is the dot level relative to the electrochemical potential of the ferromagnet,  tunable by the gate voltage.  Furthermore, $U$ is the local Coulomb interaction in the dot,  $p$ is the spin polarization of the ferromagnet, and   $W$ denotes the half-width of the conduction band.
Notably, the exchange field, plotted in \Fig{fig:2}a, depends on the gate voltage in an antisymmetric way, $B(\varepsilon)=-B(-\varepsilon-U)$, reversing its sign at the symmetry point $\varepsilon = -U/2$.
The electron- ($B_0$) and hole-type ($B_2$) fluctuations cancel there, as experimentally observed by Hauptmann~\emph{et~al.}~\cite{Hauptmann_NaturePhys.4/2008}.
This is a generic feature of interacting quantum-dot spin-valves~\cite{Sindel_Phys.Rev.B76/2007}
if spin-polarization effects of the ferromagnets dominate~\cite{Gaass_Phys.Rev.Lett.107/2011}.
Approaching the symmetry point, deep in the Coulomb blockade, processes of higher order in $\Gamma$ become increasingly important.
These are responsible for inelastic tunneling, as well as the Kondo effect, both being primary experimental tools in atomic-scale spin detection~\cite{Madhavan_Science280/1998,Heinrich_Science306/2004,Meier_Science320/2008} and manipulation~\cite{Loth_NaturePhys.6/2010}.
The interplay of such processes  with the exchange field $B(\varepsilon)$ has been analyzed~\cite{Martinek_Phys.Rev.Lett.91/2003_127203,Sindel_Phys.Rev.B76/2007}
and experimentally demonstrated for $S=1/2$ molecular~\cite{Pasupathy_Science306/2004} and carbon-nanotube~\cite{Hauptmann_NaturePhys.4/2008,Gaass_Phys.Rev.Lett.107/2011} quantum dots.

However, for high-spin quantum dots these processes result in a drastically different situation as we now explain.
For our $S=1$ quantum dot example processes of the order $\Gamma^2$, apart from inessential renormalization of $B(\varepsilon)$,  generate an  additional \emph{spintronic anisotropy term} of the same form as the second term of \Eq{eq:heff}.
The result for the \emph{quadrupolar exchange field} $D(\varepsilon)$ (see Methods) is plotted in \Fig{fig:2}a.
It takes a simple, approximate form in the regime $|\varepsilon+U/2| \ll U/2$
when we neglect the excited singlet state at energy $|K|$ and assume a large band width $W \gg U  \gg T \gg \Gamma$ [as was also done in deriving \Eq{eq:B}]:
    \begin{align}
    D(\varepsilon) = - B_0(\varepsilon) \frac{\partial B_0(\varepsilon)}{\partial\varepsilon}
    - B_2(\varepsilon) \frac{\partial B_2(\varepsilon)}{\partial(-\varepsilon)}.
    \label{eq:D}
    \end{align}
The corresponding term in \Eq{eq:heff} generates a quadrupolar splitting. We find that in the Coulomb blockade regime this $D$ parameter is negative, i.e. the energy of the triplet \emph{axial} spin states $\ket{S_z=\pm1}$ is lowered relative to the \emph{planar} spin state $\ket{S_z=0}$, as in \Fig{fig:1}b.
This is entirely analogous to the uniaxial spin anisotropy typical to  magnetic adatoms or molecules, cf. \Fig{fig:1}a.
However, this  anisotropy is induced by the proximity of the ferromagnet and displays the characteristic properties of a \emph{spintronic} exchange field: it is electrically tunable by the gate voltage as shown in~\Fig{fig:2}a, and scales as $D\propto p^2\Gamma^2$.
It is thus enhanced with increasing tunnel coupling $\Gamma$, similar to the Kondo effect (see below), but in contrast, it is also enhanced with increasing spin polarization $p$,
which suppresses the Kondo effect.

Importantly, \Fig{fig:2}a demonstrates that the quadrupolar field  is a symmetric function of the gate voltage, $D(\varepsilon)=D(-\varepsilon-U)$, and therefore does not necessarily vanish at the symmetry point [the electron and hole contributions to \Eq{eq:D} \emph{add up}, unlike in \Eq{eq:B}].
Expanding $B(\varepsilon)$ and $D(\varepsilon)$  linearly  around the symmetry point,
    \begin{align}
    B(\varepsilon) & \approx   - \frac{2}{\pi} p \Gamma \frac{\varepsilon+{U}/{2}}{U}
    &  & &\mathrm{ (linear),}
    \label{eq:Bstar}
    \\
    D(\varepsilon) & \approx   -\frac{1}{\pi^2}  \frac{(p \Gamma)^2 }{U} \text{ln} \frac{2 W}{ U}
    :=  D^{*}&
    & & \mathrm{ (constant),}
    \label{eq:Dstar}
    \end{align}
we thus obtain an all-spintronic superparamagnet described by \Eq{eq:heff} with a constant anisotropy $D^{*}$ and a magnetic field that is linearly tunable by  the gate voltage through $\varepsilon$, see \Fig{fig:2}b.
Close to the symmetry point, the quadrupolar field dominates over the dipolar one in the gate voltage range
$|\varepsilon+{U}/{2}|
\ll
\delta\varepsilon :=
p\Gamma \text{ln}(2W/U)/(2\pi)
$
proportional to the width $\Gamma$ of the low-temperature Coulomb peaks.
Accordingly, high-spin quantum dot spin valves exhibit  a tunable interplay of spintronics and nanomagnetism that is not possible for low-spin quantum dots.
This, in turn, opens the possibility for fast all-electric operations involving the spin, which are challenging for adatoms and single-molecule magnets.
%

%================================================================================================
\section*{Spectral signatures of spintronic quadrupolar splittings}

Based on the perturbative  results \eq{eq:Bstar}-\eq{eq:Dstar} we expect
a clear experimental \emph{transport} signature of the spintronic quadrupolar field
for strong tunnel coupling $\Gamma$.
The above considerations are readily extended to the case of a junction of two ferromagnets with voltage bias $V_\textrm{b}$ and parallel polarizations (see Methods).
The spintronic fields $B$ and $D$ now also acquire a dependence on the bias voltage, which is, however, negligible in the situation under discussion (see \suppinfo).
In order to address the strong tunneling regime and to better estimate the achievable magnitude of $D$ we  calculate the equilibrium, spin-resolved local density of states (LDOS)  using the exact density-matrix numerical renormalization group (DM-NRG) method.
This allows us to compute transport characteristics, while
(i) including the singlet excited state at finite energy $|K|$ that we neglected so far,
and (ii) taking into account  the Kondo effect,
which also gains importance with increasing $\Gamma$ at low temperature.

To set the stage, we show in \Fig{fig:3}a the result for a spin $S=1/2$ quantum-dot spin valve model (obtained by setting $K=0$), successfully used to analyse  the spectroscopy of the dipolar exchange field~\cite{Hauptmann_NaturePhys.4/2008,Gaass_Phys.Rev.Lett.107/2011}.
Unlike for non-magnetic electrodes,
a Kondo peak forms \emph{only} at the symmetry point where the exchange field $B$ induced by the ferromagnets vanishes~\cite{Choi_Phys.Rev.Lett.92/2004,Martinek_Phys.Rev.Lett.91/2003_127203,Sindel_Phys.Rev.B76/2007}.
The finite-temperature precursor  of the Kondo peak in \Fig{fig:3}a
displays the measured, gate-voltage dependent splitting~\cite{Hauptmann_NaturePhys.4/2008}.

For a high-spin $S=1$ ground state (i.e. $K<0$), instead of a peak, we find in  \Fig{fig:3}b   a pronounced gap, which linearly increases as the gate voltage is detuned from the symmetry point.
This indicates a definite spin excitation, even close to the symmetry point where the influence of a dipolar exchange field $B$ on the high-spin quantum dot is negligible.
As illustrated in \Fig{fig:2}b, the observed excitation  as a function of $\varepsilon+U/2$ is the telltale signature of uniaxial spin anisotropy of the type predicted by equations~\eq{eq:heff} and \eq{eq:D}.
By relating the gate voltage through \Eq{eq:Bstar} to the magnetic field, this signature is seen to coincide with that of  \emph{intrinsic} anisotropy  discussed theoretically and observed experimentally in the transport through real magnetic adatoms and molecules~\cite{Jo_NanoLett.6/2006,Otte_NaturePhys.4/2008,Zyazin_NanoLett.10/2010}.
The failure to close the gap  at the symmetry point in \Fig{fig:3}b  corresponds to a so-called `zero-field splitting'~\cite{Gatteschi_book} of such systems.
During the excitation the spin transits from the lowest axial state $\ket{S_z=\pm1}$
into the planar state $\ket{S_z=0}$, see \Fig{fig:2}b.
Direct transitions between the axial spin levels are spin-forbidden and do not show up in \Fig{fig:3}b.
Importantly, the different energy units on the left and right $\omega$-axis in Figs.~\ref{fig:3}c,d  reveal that the DM-NRG gap is indeed of spintronic origin: it scales with $\Gamma$ and $p$ as predicted by equations~\eq{eq:Bstar}-\eq{eq:Dstar} for $|B| \gg |D|$ and $|B| \ll |D|$, respectively.
 We note that only for much stronger coupling the Kondo effect in \Fig{fig:3}b reinstates the characteristics similar to that in \Fig{fig:3}a. Finally, the quadrupolar gap can also be extracted from the temperature dependence of the transport quantities, see supplementary figure~S-6.

%%%%%%%%%%%%%%%%%%%%%%%%%%%%%%%%%%%%%%%%%%%%%%%%%%%%%%%%%%%%%%%%%%%%%%%%%%%%%%%%%%%%%%%%%%%%%%%
\begin{figure}[t]
   \includegraphics{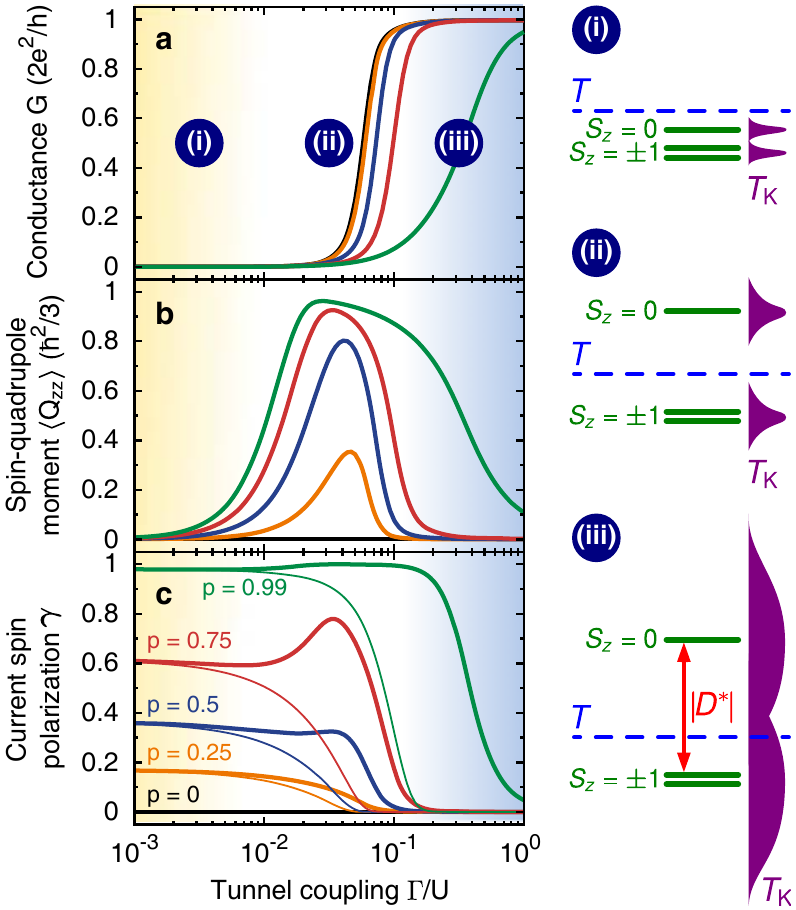}
  \caption{
    \textbf{Spin filtering}. (DM-NRG)
    \textbf{a}, Linear conductance $G = G_\uparrow+G_\downarrow$,
    \textbf{b}, average spin-quadrupole moment $\langle \hat{Q}_{zz}\rangle$ (normalized to its maximal value),
    \textbf{c}, linear response spin polarization of transported electrons, $\gamma=(G_\uparrow-G_\downarrow)/(G_\uparrow+G_\downarrow)$,
    shown as a function of the tunnel coupling $\Gamma$ for various values of spin polarization $p$ of the ferromagnets at the symmetry point $\varepsilon=-U/2$.
    Solid (thin) lines correspond to the spin $S=1$ ($S=1/2$) quantum dot.
    Side diagrams (i)-(iii) illustrate schematically the competition of the energy scales of temperature $T$ (dashed lines), spintronic anisotropy $D^\ast(\Gamma)$ (red) and Kondo spin scattering $T_\textrm{K}\big(\Gamma,D^\ast(\Gamma)\big)$ (purple), as found by DM-NRG (see \suppinfo). Due to the  $\Gamma$-dependence, starting from thermal regime (i), first $|D^\ast(\Gamma)|$ rises faster, (ii), and ultimately the Kondo effect catches up, (iii).
    Parameters are as in~\Fig{fig:3}, except for
    $T=7.5\cdot10^{-6}U \approx 9$ mK, which is smaller than the temperature needed to resolve the  maximally achievable gap $2|D^{*}(\Gamma)|$ in transport spectra of \Fig{fig:3}b (i.e. for $\Gamma$ values just before the Kondo effect sets in).
  }
  \label{fig:4}
\end{figure}
%================================================================================================
By attaching the ferromagnets we have thus obtained an artificial molecular magnet,
whose quadrupolar field $D$ is strong enough to suppress the Kondo effect in a wide range of parameters, see \suppinfo. This is expected from the analysis\cite{Romeike_Phys.Rev.Lett.96/2006,Zitko_Phys.Rev.B78/2008,Misiorny_Phys.Rev.B84/2011} of the destructive effect of \emph{intrinsic} uniaxial anisotropy on the Kondo resonance~\cite{Otte_NaturePhys.4/2008,Zyazin_NanoLett.10/2010}. Notably, here \emph{both} these effects are generated by $\Gamma$, cf.~\Fig{fig:4}.
This has important experimental implications:
not only does spintronic anisotropy modify  an existing, intrinsic anisotropy barrier (induced by spin-orbit interaction)~\cite{Misiorny_Phys.Rev.B86/2012},
but it can even create such a barrier from scratch (without spin-orbit coupling),
thus \emph{generating} the entire observed spin-excitation spectrum.
Consequently, care must be taken when using spin-polarized electrodes to search for signatures of intrinsic superparamagnetism:
tunneling-renormalization effects, observed for electronic excitations~\cite{Grove-Rasmussen_Phys.Rev.Lett.108/2012}, also affect the quantitative extraction of intrinsic anisotropy parameters,
in particular when the spin polarization becomes significant, as desired in spintronics, and the tunnel coupling becomes strong.
So far, the parameters were chosen to enable a comparison of the results obtained by the complementary  methods used. Having established the scaling~\eq{eq:Dstar}, we can now
give a conservative estimate for the spintronic anisotropy barrier: it
can be as large as 0.04~meV, i.e. the gap $2|D^{*}|$ is clearly
resolvable at $T\sim 230$ mK, see \suppinfo. This value is
comparable with $|D|$ in the range of tens of $\mu$eV for
single-molecule magnets~\cite{Gatteschi_book,Zyazin_NanoLett.10/2010,Kahle_NanoLett.12/2012},
while $|D|$ can reach up to few meV for some adatoms~\cite{Brune_Surf.Sci.603/2009,Hirjibehedin_Science317/2007}.

\vspace{-0.5cm}
\section*{Enhanced spin filtering}

Finally, the spintronic quadrupolar exchange field  can also be used to enhance the linear-response spin filtering of electrons transported  through a quantum dot with \emph{zero dipolar exchange field and average spin}.
\Fig{fig:4}a presents the linear conductance (DM-NRG) as function of the tunnel coupling $\Gamma$ for fixed, intermediate temperature  at the symmetry point where $\braket{\veco{S}}=0$
(see \suppinfo).
We observe that with increasing spin polarization $p$ of the ferromagnets, the   value of $\Gamma$ for which the Kondo unitary conductance is reached strongly increases.
This shift is caused by the spintronic anisotropy, evidenced by the finite quadrupolarization $\braket{\hat{Q}_{zz}}>0$  in \Fig{fig:4}b, which arises when $|D^\ast(\Gamma)| \gg T$ -- well before the Kondo effect sets in. Interestingly, the conductance spin polarization  in \Fig{fig:4}c shows a corresponding peak that develops into a 100\%-plateau for $p$ close to, yet still less than 1.
But even for 50\% polarized ferromagnets ($p=0.5$) the conductance spin polarization can be almost doubled if the temperature is further reduced, see supplementary figure~S-8.

This amplification of the conductance spin polarization is a hallmark of the exchange quadrupolar field:
in a broad, intermediate regime of $\Gamma$-values the spintronic anisotropy  gap $|D^{*}| \gg T$ makes the planar spin state $\ket{S_z=0}$ inaccessible at low energy, as depicted in \Fig{fig:4}(ii).
In consequence, tunneling induced spin-flip processes between the axial states $\ket{S_z=\pm 1}$ (Kondo effect) are strongly suppressed,  greatly enhancing the conductance spin polarization, see \suppinfo.
Such a behaviour is absent in the results for $S=1/2$ shown for comparison as thin curves in \Fig{fig:4}c:
in this case, spin reversal does not involve a planar state.

This enhanced spin-filtering effect illustrates a promising synergy of spintronics and nanomagnetism.
The spintronic anisotropy may offer new possibilities for combining tools and insights from these two research fields.
As we show in \suppinfo, one may envision to utilize the quadrupolar field as a new means
for fast, \emph{electrical} control of
 nanomagnetic memory cells, allowing even  for on-demand bistability.
 Furthermore, in contrast to intrinsic anisotropy,
 the spintronic anisotropy can also be turned off \emph{magnetically} by switching the spin-valve to the antiparallel configuration.
Finally, it is interesting to ask whether spintronic anisotropy may be understood in a broader context: just as the dipolar exchange field term in \Eq{eq:heff}, expressing the spin torque, can be seen as a part of the spin(-dipole) current,
so, perhaps, the quadrupolar exchange field term may relate to a spin-quadrupole current~\cite{Baumgartel_Phys.Rev.Lett.107/2011}.

%================================================================================================
%  Reduce fontsize for the rest % {fontsize}{spacingsize}
\fontsize{8}{9}\selectfont
%

%================================================================================================
\section*{Methods}
The model of a  high-spin quantum-dot spin valve in~\Fig{fig:1}c consists of a single spin-degenerate orbital level, tunnel-coupled to two ferromagnets, and involving the local \emph{isotropic} Heisenberg exchange coupling $K<0$ to an immobile $S=1/2$ impurity. Electron tunneling through the junctions is assumed to be symmetric and spin conserving.

\vspace{-0.5cm}
\subsection*{Real-time diagrammatic (RTD) perturbation theory}
By integrating out the ferromagnets held at different equilibria we obtain
a stationary quantum kinetic equation for the reduced density operator $\rho$,
$\dot{\rho}(t) = 0 = -i [\mathcal{H}_\textrm{dot},\rho] -i \Sigma \rho$, where $\Sigma$ is the zero-frequency kernel expanded formally in powers of $\Gamma$.
We rewrite $\Sigma \rho = [\mathcal{H}_\textrm{eff},\rho] + \Sigma' \rho$
where $\mathcal{H}_\textrm{eff}$ is the renormalization of the Hamiltonian $\mathcal{H}_\textrm{dot}$
including both orders $\Gamma$ and $\Gamma^2$
which can be identified diagrammatically and then calculated.
See Sec.~IIA of \suppinfo\ for a more detailed account.
In the Coulomb blockade regime $\mathcal{H}_\textrm{dot}$ reduces to a trivial constant and $\mathcal{H}_\textrm{eff}$ is given by \Eq{eq:heff},
where $D$ for a \emph{single} electrode has the form
    \begin{equation}
    \begin{aligned}
    D = \ &\frac{\Gamma^2 p^2}{4}
    \text{Re}
    \int_{-W}^{W} \frac{\mathrm{d} \omega_1}{\pi}
    \frac{\mathrm{d} \omega_2}{\pi} \frac{ [1-f(\omega_1)] f(\omega_2) }{{\omega_2-\omega_1+i0}} \\ &
    \left(  \frac{1}{\varepsilon-\omega_1+i0}
    +
    \frac{1}{\omega_2 - (\varepsilon+U)+i0} \right)^2
    .
    \end{aligned}
    \label{eq:Dbig}
    \end{equation}
This was used to plot \Fig{fig:2}a.
In the limit $U \ll W$, the expression  reduces to the analytic result~\eq{eq:D},
from which \Eq{eq:Dstar} follows.
As mentioned in the text, one can extend the results to the case of two electrodes $r=L,R$, with respective electrochemical potentials $\mu_{L/R}=\pm V_\textrm{b}/2$:
 replace in \Eq{eq:B} $f(\omega) \rightarrow
f_r(\omega)=\big[\textrm{e}^{(\omega-\mu_r)/T)}+1\big]^{-1}$ and sum over $r$.
Equation~\eq{eq:D} remains valid as long as $V_\textrm{b} \ll U$.
Finally, we recall that the dipolar exchange field in order $\Gamma$ vanishes at the symmetry point $\varepsilon=-U/2$, cf. \Eq{eq:Bstar} and \Fig{fig:2}a.
This also holds for irrelevant higher-order corrections in $\Gamma$ to $B$,
as the zero average spin, obtained from DM-NRG calculations at the symmetry point, demonstrates for any values of $p$ and $\Gamma$.

\vspace{-0.5cm}
\subsection*{Density-matrix numerical renormalization group}
We use the \emph{Flexible} DM-NRG~\cite{Toth_Phys.Rev.B78/2008} approach  to  numerically calculate the spin-resolved equilibrium spectral function $a_\sigma(\omega)$ [$\sigma=\uparrow,\downarrow$], referred to also as LDOS, of the  orbital level for  parallel polarization of the ferromagnets.
From this we compute the spin-resolved linear-response conductance
    \begin{equation}
    G_{\sigma}=\frac{2e^2}{h}\pi  \frac{\Gamma(1+\eta_\sigma p)}{2}\int\textrm{d}\omega \left[-\frac{\partial f(\omega)}{\partial\omega}\right]
   a_{\sigma}(\omega),
   \label{eq:Gsigma}
    \end{equation}
where $\eta_{\uparrow(\downarrow)}=\pm1$.
Moreover, by approximating $\textrm{d}I/\textrm{d}V_\textrm{b} \propto A(\omega)$ at low $T$ and low, but finite bias voltage $V_\textrm{b}=\omega$ with the symmetrized, dimensionless equilibrium spectral function  $A(\omega)=\pi\Gamma\sum_\sigma(1+\eta_\sigma p)\big[a_\sigma(\omega)+a_\sigma(-\omega)\big]/2$,
we can infer useful qualitative conclusions about the differential conductance, taking into account known limitations of this approximation.
Further details can be found in Sec.~IIB of \mbox{\suppinfo.}
%================================================================================================
%\bibliographystyle{naturemag}
%\bibliography{Bib_NaturePhys}

\vspace{-0.5cm}
%================================================================================================
\acknowledgments
We acknowledge stimulating discussions with J. Barna\'{s}, A. Cornia, S. Das,
 J. K\"onig, S.~J. van der Molen, J. Splettstoesser, I. Weymann, and H. van der Zant.
The use of the SPINLAB computational facility and the open access Budapest flexible DM-NRG code~\cite{Toth_Phys.Rev.B78/2008} (\url{http://www.phy.bme.hu/~dmnrg/}) is kindly acknowledged.
We acknowledge the financial support from the  DFG (FOR 912), the Foundation for Polish Science (M.M.) and the Alexander von Humboldt Foundation (M.M.).
Correspondence and requests for materials should be addressed to M.M..
Supplementary Information accompanies this paper on \url{www.nature.com/naturephysics}.
%================================================================================================
\vspace{-0.5cm}
\subsection*{Author contributions}
M.W. conceived the idea.
M.H. and M.M. performed the analytic and numerical calculations, respectively.
M.H. provided M.M. and M.W. with fitting formulas for the physical analysis of DM-NRG results.
M.M. prepared the initial manuscript.
All authors contributed in writing the manuscript.
%================================================================================================
\vspace{-0.5cm}
\subsection*{Competing financial interests}
The authors declare no competing financial interests.

\onecolumngrid
\clearpage



\section*{Supplementary material}

# Supplementary Information for
## *'Spintronic magnetic anisotropy'*

Maciej Misiorny,[1, 2, 3] Michael Hell,[1, 2] and Maarten R. Wegewijs[1, 2, 4]

[1]*Peter Grünberg Institut, Forschungszentrum Jülich, 52425 Jülich, Germany*
[2]*JARA – Fundamentals of Future Information Technology*
[3]*Faculty of Physics, Adam Mickiewicz University, 61-614 Poznań, Poland*
[4]*Institute for Theory of Statistical Physics,*
*RWTH Aachen, 52056 Aachen, Germany*

## CONTENTS

In this Supplementary Information we provide a precise formulation of the model used in the main text (Sec. I), then discuss the application of the employed methods specifically to this model (Sec. II), expanding the 'Methods' section of the main article. Finally, in Sec. III we provide an extensive discussion of claims and results in the order they appear in the main article. Within the Supplementary Information references are numbered as, e.g., equation (S-1) and Figure S-1, whereas regular numbers, e.g., equation (1) and Figure 1, refer to the main article.

## I. MODEL

In the main article we illustrate the spintronic generation of anisotropy by analysing the transport properties of a high-spin quantum dot embedded between two ferromagnetic electrodes. The full Hamiltonian of the system under consideration reads as $\mathcal{H} = \mathcal{H}_{\text{dot}} + \mathcal{H}_{\text{el}} + \mathcal{H}_{\text{tun}}$.

The high-spin quantum dot, described by

$$\mathcal{H}_{\text{dot}} = \varepsilon \hat{n} + U \hat{n}_{\uparrow} \hat{n}_{\downarrow} + K \hat{\mathbf{s}} \cdot \hat{\mathbf{S}}_{\text{imp}}, \tag{S-1}$$

is a composite system consisting of a single-level quantum dot coupled to a immobile magnetic spin-1/2 impurity with ferromagnetic **Heisenberg** interaction $K < 0$[1,2]. Here, $\varepsilon$ denotes the gate-tunable energy of a single electron occupying the dot, while $U$ stands for the Coulomb energy of two electrons with opposite spins when the dot is doubly occupied. Moreover, $\hat{n} = \sum_{\sigma} \hat{d}_{\sigma}^{\dagger} \hat{d}_{\sigma}$ represents the dot's charge operator, with $\hat{d}_{\sigma}^{\dagger}$ ($\hat{d}_{\sigma}$) being the creation (annihilation) operator of a spin-$\sigma$ electron ($\sigma = \uparrow, \downarrow$) in the dot. Consequently, whenever a single electron resides in the dot,

its spin (characterised by the operator $\hat{s}$) forms a $S = 1$ ground state with the impurity spin with the operator $\hat{S}_{imp}$.

Such a quantum dot model describes various experimental systems. For instance, it arises in semiconductor heterostructures, in asymmetrically gated a double quantum dots[3] or multilevel quantum dots[4,5]. It accounts for the physics observed in gold nanoparticles coupled to a magnetic impurity[6] and also a molecule hyperfine-coupled to an internal, single nuclear spin[7]. Furthermore, in nanojunctions, it was demonstrated that the sign and magnitude of $K$ can be electrically tuned *in situ*[8–10]. The crucial feature that the Heisenberg interaction in this model accounts for is the gap separating the high-spin state from the excited states, irrespective of their nature (i.e., with either smaller or larger spin). In transport through an endofullerene molecule the high-spin ground state was shown to be separated by 0.8 meV from the excitations[11], in this case caused by antiferromagnetic Heisenberg interaction. Molecules with even larger gaps abound in the chemistry of molecular magnets, where the main synthetic challenge is to generate a high anisotropy $D$ in high-spin molecules with a large exchange gap. In these systems gaps are caused by a non-trivial competition between several types of exchange interactions and can be several meV large, e.g. 6 meV for the well-characterized $Fe_4$ class single-molecule magnets[12]. Using a 'bad' molecular magnet (i.e., with a large exchange splitting as in Ref. [12], but with negligible $D$) as a quantum dot in the setup suggested in this article, can thus also be a good starting point for spintronic generation of anisotropy. With this broad class of high-spin quantum dot systems in mind we set in the numerical computations $K = -1$ meV. We note that larger values of $K$ also facilitate the comparison with the analytical real-time diagrammatic calculations which are tractable only for $|K| \to \infty$, see Sec. II A Finally, in Fig. S-7 and its caption we show that the main conclusions remain qualitatively valid also for smaller magnitudes of the Heisenberg interaction $K$.

The ferromagnetic metallic electrodes are described by non-interacting itinerant electrons with energy dispersion $\varepsilon_{k\sigma}$, $\mathcal{H}_{el} = \sum_{rk\sigma} \varepsilon_{k\sigma} c_{k\sigma}^{r\dagger} c_{k\sigma}^{r}$ with $c_{k\sigma}^{r\dagger}$ ($c_{k\sigma}^{r}$) being the creation (annihilation) operator for the $r$th electrode ($r = L, R$), and $k$ denoting an orbital quantum number. For simplicity, we assume that the electrodes are identical, have *parallel* spin-polarization vectors, and are described by a constant spin-dependent density of states (DOS), $\sum_{k} \delta(\varepsilon_{k\sigma} - \omega) \approx \nu_{\sigma}$, for $|\omega| \leq W$, where $W$ is the half-width of the conduction band. We emphasize that here and above all spin quantum numbers refer to the $z$-axis taken to in the spin-polarization direction.

Next, electron tunneling processes between the dot and electrodes are accounted for in a standard way[13] by

$$\mathcal{H}_{tun} = \sum_{rk\sigma} t_{\sigma}^{r} d_{\sigma}^{\dagger} c_{k\sigma}^{r} + \text{H.c.},$$ (S-2)

where $t_{\sigma}^{r}$ stands for the tunnel matrix element between the dot and the $r$th electrode. We assume that the junctions are symmetric ($t_{\sigma}^{L} = t_{\sigma}^{R} = t_{\sigma}$). In both methods applied below, the overall effect of the ferromagnetic electrode on the dot can then be included through the hybridization function $\Gamma_{\sigma}^{L} = \Gamma_{\sigma}^{R} \equiv \Gamma_{\sigma}$,

$$\Gamma_{\sigma} = \pi \nu_{\sigma} |t_{\sigma}|^2 = \frac{\Gamma}{2}(1 + \eta_{\sigma} p), \quad \Gamma = \Gamma_{\uparrow} + \Gamma_{\downarrow}, \quad p = \frac{\Gamma_{\uparrow} - \Gamma_{\downarrow}}{\Gamma_{\uparrow} + \Gamma_{\downarrow}}$$ (S-3)

where here and below $\eta_{\uparrow(\downarrow)} = \pm 1$, and $p$ is the spin polarization parameter of the electrodes.

Finally, we emphasize the motivation for choosing the model of a single-level quantum dot ferromagnetically interacting with a spin-one-half, immobile impurity: it is conceptually simple and allows for an analytical treatment, which complements the rigorous numerical calculations by providing detailed insight into the novel idea of the quadrupolar field. The model is representative of a class of systems which share the following crucial ingredients: (i) a high-spin ground state separated by a finite gap from further excitations, (ii) Coulomb blockade stabilizing the charge state with the high spin, and (iii) tunnel-coupling to ferromagnets. In our high-spin model the excitation gap is equal to $|K|$ which is provided by the Heisenberg interaction of the quantum dot with the impurity spin. At the end of Sec. II A we are in the position to formulate a general physical argument for the generic nature of this model. Then in Sec. III E, we check by explicit numerical calculations that our predictions remain valid not only for models in which the immobile impurity spin is larger than the minimal value $S_{imp} = 1/2$ [see Fig. S-9(a)-(b) for $S_{imp} = 1, 3/2, 2$], but also for the case in which the impurity spin is subject to charge fluctuations [see Fig. S-9(c)-(d)].

## II. METHODS

### A. Real-time diagrammatic (RTD) technique

The real-time diagrammatic (RTD) technique provides an analytic description of the high-spin quantum dot. This approach is based on a kinetic/generalized master equation for its reduced density operator $\rho$, incorporating the effect of the environment. This is accomplished by integrating out the ferromagnets (held at different equilibria with the same temperature $T$, but different electrochemical potentials $\mu_r$). In the stationary limit we obtain the quantum master equation $\dot{\rho}(t) = 0 = -i[\mathcal{H}_{\text{dot}}, \rho] - i\Sigma\rho$ with the zero-frequency self-energy (or time-evolution kernel) $\Sigma$. The self-energy superoperator $\Sigma$ is formally expanded in powers of $\Gamma$ using the standard real-time diagrammatic technique (RTD)[14–16] (which is equivalent to the Nakajima-Zwanzig approach[17]).

The calculation of $\Sigma$ up to the *next-to-leading* order ($\Gamma^2$) is carried out most efficiently using a Liouville-space formulation[16] in the form recently given in Ref. [18], which exploits the advantages of second-quantization in Liouville-Fock space (super-fermions). We have further adapted this approach to spin-polarized electrodes, obtaining a formulation which at every step of the calculation is explicitly 'coordinate-free', i.e. invariant under changes of both the real-space coordinate system and the spin-quantization axis. For a simple spin-valve model (i.e. without a quantum dot) this is described in detail in Ref. [19], and a full account for quantum dot spin-valves will be published in Ref. [20].

In a second step, we developed a projection technique, which uniquely extracts the part of *any* self-energy superoperator $\Sigma$ that generates a Hamiltonian evolution on the density matrix $\rho$, that is, we decompose

$$\Sigma\rho = [\mathcal{H}_{\text{eff}}, \rho]_- + \Sigma'\rho. \tag{S-4}$$

Here, $\Sigma'$ only incorporates the dissipative, non-Hamiltonian evolution induced by the bath, while the pure Hamiltonian part of the bath-induced time evolution is solely generated by the commutator with the self-adjoint effective Hamiltonian $\mathcal{H}_{\text{eff}}$. The latter can be extracted from the matrix elements $\Sigma_{AB} = \text{tr}\{A^\dagger(\Sigma B)\}$ of the full self-energy by

$$\mathcal{H}_{\text{eff}} = \frac{1}{6}\sum_{A,B\in\mathcal{T}}\Sigma_{AB}\,[A,B]_- + \text{other terms}. \tag{S-5}$$

In the equation above, the first term is the projection of the effective Hamiltonian on the triplet space and the sum runs over a complete set of operators for the traceless triplet subspace of the quantum dot Liouville space[20]: $\mathcal{T} = \{\frac{1}{\sqrt{2}}\hat{S}_i^t\}_{i=x,y,z}\cup\{\hat{Q}_{ij}^t\}_{i,j=x,y,z}$. The superscript $t$ denotes the projection $P^t$ onto triplet states, $\hat{A}^t = P^t\hat{A}P^t$, of the spin-dipole operators $\hat{S}_i$ and spin-quadrupole operators

$$\hat{Q}_{ij} = \frac{1}{2}\left(\hat{S}_i\hat{S}_j + \hat{S}_j\hat{S}_i\right) - \frac{1}{3}\hat{\mathbf{S}}^2\delta_{ij}, \tag{S-6}$$

where $i,j = x,y,z$ are their Cartesian components. The 'other terms' in equation (S-5) represent the effective Hamiltonian evolution of the other charge states (empty or doubly occupied level). Deep inside the Coulomb blockade regime, which is of interest here, only one electron occupies the tunnel-coupled orbital and contributions from other charge states are negligible. Furthermore, we excluded the high-lying singlet excitation for the sake of simplicity. We note that the above goes beyond the standard procedure for calculating renormalization of the subsystem by Lamb-shifts which only account for the leading order. In spintronics context, we thus extend Refs. [2] and [21–22] by including not only order $\Gamma$ but also order $\Gamma^2$ terms. From a general perspective, the quadrupolar exchange field is a striking example where the next-to-leading order renormalization of a subsystem Hamiltonian provides the crucial physics (that even dominates the Kondo effect in large parameter regimes).

In a third step, we recast equation (S-5) into the form given in the main article as equation (1),

$$P^t\mathcal{H}_{\text{eff}}P^t = B\hat{S}_z^t + D\hat{Q}_{zz}^t. \tag{S-7}$$

Some elaboration on the commutators in equation (S-5) shows that the number of the supermatrix elements $\Sigma_{AB}$ needed to compute either of the exchange fields $B$ and $D$ can be greatly reduced.

This enables a targeted calculation of the exchange fields without computing the *full* self-energy $\Sigma$. In particular, the quadrupolar field of interest here reads

$$D = \frac{1}{12} \sum_{i,j,k} \left( \Sigma_{\hat{Q}^t_{ij} S^t_k} \mathrm{tr}\left\{ \hat{Q}^t_{ij} \left[ \hat{Q}^t_{ij}, \hat{S}^t_k \right]_{-} \right\} + \hat{Q}^t_{ij} \leftrightarrow \hat{S}^t_k \right). \tag{S-8}$$

We find the specific self-energy supermatrix element $\Sigma_{\hat{Q}^t_{ij} \hat{S}^t_k}$ by applying the real-time diagrammatic technique up to the order $\Gamma^2$ and obtain

$$D = \frac{\Gamma^2 p^2}{4} \mathrm{Re} \int_{-W}^{+W} \frac{\mathrm{d}\omega_1}{\pi} \int_{-W}^{+W} \frac{\mathrm{d}\omega_2}{\pi} \frac{\left[ 1 - f(\omega_1) \right] f(\omega_2)}{\omega_2 - \omega_1 + i0} \left( \frac{1}{\varepsilon - \omega_1 + i0} + \frac{1}{\omega_2 - (\varepsilon + U) + i0} \right)^2, \tag{S-9}$$

which is equation (6) in the 'Method' section of the main article. Here, the spin-dependence enters *via* the spin-dependence of DOS[23], i.e. $t_\sigma = t$. The DM-NRG calculations confirm that the quadrupolar exchange field scales quadratically with the tunnel coupling $\Gamma$ and the spin polarization $p$ of the leads at the symmetry point, $\varepsilon = -U/2$, exactly as predicted by equation (S-9), see Sec. III B 2 for further discussion. The dependence of $D^* = D(\varepsilon = -U/2)$ on the interaction energy $U$ for low $T \ll U$ can be estimated by substituting $2\omega_1/U = w_1$ and $2\omega_1/U = -w_1$ in (S-9), and approximating the Fermi functions by step functions. This yields

$$D^* = -\frac{\Gamma^2 p^2}{U} C \left( \frac{2W}{U} \right) \tag{S-10}$$

where to leading order in $U/W$, we reproduce equation (5) of the main article, since

$$
\begin{aligned}
C(\Lambda) &= \frac{1}{2} \mathrm{Re} \int_0^\Lambda \frac{\mathrm{d}w_1}{\pi} \int_0^\Lambda \frac{\mathrm{d}w_2}{\pi} \frac{1}{w_1 + w_2 + i0} \left( \frac{1}{1 + w_1 + i0} + \frac{1}{1 + w_2 + i0} \right)^2 \\
&= \frac{1}{\pi^2} \left[ \ln(\Lambda) + O\left(1\right) \right].
\end{aligned}
\tag{S-11}
$$

The numerical evaluation of (S-9) for all level positions $\varepsilon$ requires a fourth step. We convert the double frequency integral into a double summation over Matsubara frequencies by first substituting $x_2 = \omega_2/T$ and $x_1 = -\omega_2/T$ and splitting the Fermi functions $f(xT) = g^+(x) + g^-(x)$ into their symmetric part $g^+(x) = 1/2$ and their antisymmetric part $g^-(x) = -\tanh(x/2)/2$, respectively. We then integrate over $x_1$ and $x_2$ using complex integration, closing the integration contour in the upper half of the complex plane. By virtue of the residue theorem, one can derive the following relation for the generic type of integrals occurring after these manipulations

$$
\begin{aligned}
&\int_{-R}^{+R} \mathrm{d}x_1 \int_{-R}^{+R} \mathrm{d}x_2 \; g^{q_1}(x_1) g^{q_2}(x_2) \frac{1}{x_i - \lambda_2 + i0} \frac{1}{x_1 + x_2 + i0} \frac{1}{x_1 - \lambda_1 + i0} \\
&= \delta_{q_1,-} \sum_{k_1,k_2}^{k_R} \left[ -4\pi^2 \delta_{q_2,-} \frac{1}{z_{k_i} - \lambda_2} \frac{1}{z_{k_1} + z_{k_2}} \frac{1}{z_{k_1} - \lambda_1} + 2\pi i \delta_{i,1} \frac{1}{z_{k_1} - \lambda_2} M^{q_2}_{k_2} \frac{1}{z_{k_1} - \lambda_1} \right] + O\left( \frac{1}{R} \right),
\end{aligned}
$$

where $q_1, q_2 = \pm$ and $z_k = i\pi(2k + 1)$ are the Matsubara frequencies, and $0 \le k \le k_R = \lceil \frac{R}{2\pi} - \frac{1}{2} \rceil$ with $\lceil x \rceil$ denoting the smallest integer that is not less than $x$. We additionally used the abbreviation

$$M^{q_2}_{k_2} = \frac{1}{2} \left[ q_2 \ln \left( \frac{z_{k_2} + iR}{z_{k_2} + R} \right) + \ln \left( \frac{z_{k_2} - R}{z_{k_2} + iR} \right) \right]. \tag{S-12}$$

The above double Matsubara sums are computed numerically, yielding the plot of equation (6) shown in Fig. 2(a) of the main article.

Finally, we explain how equation (6) reduces to the analytic result (3) in the large bandwidth limit $W \gg U$, as mentioned in the 'Method' section of the article. For this purpose, we expand the square of the bracketed expression in equation (S-9). In the vicinity of the symmetry point $\varepsilon = -U/2$, the two cross terms in this expansion are only large if at the same time $\omega_1 = -U/2$ and $\omega_2 = +U/2$, at which point, however, the prefactor $(\omega_2 - \omega_1 + i0)^{-1}$ is small. The remaining two terms

contain the square of each of the propagators in the bracketed expression in equation (S-9) and become maximal if either $\omega_1 = -U/2$, without restricting $\omega_2$, or $\omega_2 = +U/2$, without restricting $\omega_1$. In either case the prefactor $(\omega_2 - \omega_1 + i0)^{-1}$ can thus become large when $\omega_1 \approx \omega_2$ and dominate over the cross-terms, which we therefore neglect.

We then evaluate the remaining integrals by replacing

$$
\begin{aligned}
\frac{1}{\varepsilon_0 - \omega_1 + i0} &\rightarrow \frac{1}{\varepsilon_0 - \omega_1 + i0} + \frac{1}{\omega_2 - \varepsilon_0 + i0} = \frac{\omega_2 - \omega_1 + i0}{(\varepsilon_0 - \omega_1 + i0)(\omega_2 - \varepsilon_0 + i0)}, \\
\frac{1}{\omega_2 - \varepsilon_2 + i0} &\rightarrow \frac{1}{\omega_2 - \varepsilon_2 + i0} + \frac{1}{\varepsilon_2 - \omega_1 + i0} = \frac{\omega_2 - \omega_1 + i0}{(\varepsilon_2 - \omega_1 + i0)(\omega_2 - \varepsilon_2 + i0)},
\end{aligned}
\tag{S-13}
$$

where $\varepsilon_0 = \varepsilon$ and $\varepsilon_2 = \varepsilon + U$. One can show that this does not change the real part of the integral that is needed. Subsequently canceling the numerator on the right hand side of (S-13) with the prefactor $(\omega_2 - \omega_1 + i0)^{-1}$ in (S-9), the double frequency integral factorizes into single frequency integrals. We use that

$$
\frac{1}{(\varepsilon_n - \omega_1 + i0)^2} = -\frac{\partial}{\partial \varepsilon_n} \frac{1}{(\varepsilon_n - \omega_1 + i0)} \quad \text{for } n = 0, 2,
\tag{S-14}
$$

and apply the Sokhotsky's formula

$$
\frac{1}{x + i0} = P\left(\frac{1}{x}\right) - i\pi\delta(x).
\tag{S-15}
$$

As a result, we obtain an approximated version for the quadrupolar field in the limit $U \ll W$:

$$
D = -\left( B_0(\varepsilon)\frac{\partial B_0(\varepsilon)}{\partial \varepsilon} + B_2(\varepsilon)\frac{\partial B_2(\varepsilon)}{\partial(-\varepsilon)} + \frac{\Gamma^2 p^2}{4} f(\varepsilon)\frac{\partial f(\varepsilon)}{\partial \varepsilon} + \frac{\Gamma^2 p^2}{4}\left[1 - f(\varepsilon)\right]\frac{\partial[1 - f(\varepsilon)]}{\partial(-\varepsilon)} \right).
\tag{S-16}
$$

At low $T \ll U$, the terms involving the derivatives of the Fermi functions are exponentially small, and thus deep in the Coulomb blockade regime they can be neglected, which leads to equation (3) of the main article. Finally, we note that the order $\Gamma^2$ corrections to the dipolar exchange field $B$ can in principle be calculated in the same way, but they are not required, see Sec. III C.

As mentioned in the main paper, one can extend the above results to the case of two electrodes $r = L, R$, with respective electrochemical potentials $\mu_{L/R} = \pm V_b/2$. For this purpose, one replaces the Fermi function as follows: $f(\omega) \rightarrow \sum_r f_r(\omega)$ with $f_r(\omega) = \left[e^{(\omega - \mu_r)/T} + 1\right]^{-1}$ in equation (2) for $B$ in the main article, whereas $f(\omega_2) \rightarrow \sum_{r_2} f_{r_2}(\omega_2)$ and $1 - f(\omega_1) \rightarrow \sum_{r_1}\left[1 - f_{r_1}(\omega_1)\right]$ in equation (S-9) for $D$. After substituting the distances to the electro-chemical potentials $\omega_i = \omega_i' + \mu_{r_i}$ in all Fermi functions we can proceed with the same steps as sketched above since the resulting shifts $\mu_{r_i}$ in the propagators can be absorbed into the energies $\varepsilon_0$ and $\varepsilon_2$.

*Generic nature of the model.* At this point we can formulate a general argument for the generic nature of the high-spin quantum dot model (S-1) and the new physics arising as compared to $S = 1/2$ quantum dots ($K = 0$). Generally, coupling a quantum dot to ferromagnetic electrodes leads to a breaking of spin symmetry on the dot. In the ground multiplet of an $S = 1$ quantum dot the possible perturbation terms arising in the effective Hamiltonian (1) are represented by $3 \times 3$ matrices. Apart from a possible trivial scalar coupling amounting to a level shift (which is zero here), these terms can be classified as a vector operator (the spin operator) coupling to the dipolar exchange field vector and a traceless rank 2 tensor operator (five independent components related to spin-quadrupole moments) coupling to the quadrupolar exchange field tensor. The only simplification that occurs for our model is due to the spin-rotational symmetry with respect to the $z$-axis of the (parallel) ferromagnet(s): this implies that only components referring to the $z$-axis, denoted by $B$ and $D$, can occur: $\mathcal{H}_{eff} = B\hat{S}_z + D\hat{Q}_{zz}$ (cf. equation (1) of the main article). For $S = 1$ this exhausts all possibilities since higher rank irreducible tensor operators are identically zero. A similar analysis can be made for the density operator: $\rho = \frac{1}{3}\hat{1} + \frac{1}{2}\langle\hat{S}_z\rangle\hat{S}_z + \langle\hat{Q}_{zz}\rangle\hat{Q}_{zz}$. The crucial difference between a spin-1/2 and spin-1 system is thus the quadrupolar degree of freedom, both in the effective Hamiltonian and in the reduced density operator describing its state. In general, a $3 \times 3$ Hermitian matrix simply requires more than four basis elements, the number required for the case of a spin-1/2. For a spin-1 the additional five basis elements form

the independent components of the quadrupole tensor operator, see also Refs. [2] and [19-20]. One thus expects on general grounds that quadrupolar terms appear (resulting in splitting and suppression of the Kondo effect, spin-filtering, etc.). Our model indeed displays these effects and is therefore sufficiently generic for a study of the new concept of spintronic anisotropy. In Sec. III E we explicitly confirm this by studying various other models.

### B. Density matrix numerical renormalization group (DM-NRG)

Using the *Flexible* DM-NRG[24-26] approach we provide a complementary, numerical description of the high-spin quantum dot. We calculate the spin-resolved equilibrium spectral function $a_\sigma(\omega)$ $[\sigma =\uparrow, \downarrow]$ of the quantum dot's orbital level, which is both coupled to the side spin *via* the Heisenberg interaction and to the ferromagnets *via* electron tunneling. We will refer to $a_\sigma(\omega)$ also as (spin-resolved) local density of states (LDOS). In the linear-response regime, the conductance of the system is given by the Wingreen-Meir formula[27,28],

$$G = \frac{2e^2}{h}\pi \sum_\sigma \frac{2\Gamma_\sigma^L \Gamma_\sigma^R}{\Gamma_\sigma^L + \Gamma_\sigma^R} \int d\omega \left[ -\frac{\partial f(\omega)}{\partial \omega} \right] a_\sigma(\omega). \tag{S-17}$$

In the equation above, the hybridization function $\Gamma_\sigma^r$ [see equation (S-3)] describes the strength of the effective coupling due to spin-dependent electron tunnelling processes between the dot and electrode $r = L, R$. Moreover, $a_\sigma(\omega) = -\frac{1}{\pi}\text{Im}\langle\langle d_\sigma | d_\sigma^\dagger \rangle\rangle_\omega^{\text{ret}}$ is the spin-dependent spectral function of the dot, with $\langle\langle d_\sigma | d_\sigma^\dagger \rangle\rangle_\omega^{\text{ret}}$ denoting the Fourier transform of the retarded Green's function $\langle\langle d_\sigma | d_\sigma^\dagger \rangle\rangle_t^{\text{ret}} = -i\theta(t)\langle\{d_\sigma(t), d_\sigma^\dagger(0)\}\rangle$, where $\{\cdot, \cdot\}$ denotes the anticommutator. This leads to the expression (7) for the spin-resolved linear-response conductance in the 'Method' section of the main article

$$G_\sigma = \frac{2e^2}{h}\pi \frac{\Gamma(1 + \eta_\sigma p)}{2} \int d\omega \left[ -\frac{\partial f(\omega)}{\partial \omega} \right] a_\sigma(\omega). \tag{S-18}$$

In order to compute the required spectral function $a_\sigma(\omega)$, Wilson's numerical renormalization-group method is used[29,30] (we use $\Lambda = 2$ and we keep 2500 states), which in general allows for addressing quantum impurity problems in an exact way. First, the model Hamiltonian is subjected to the canonical transformation[31] $c_{k\sigma}^{e(o)} = \mathcal{T}_\sigma^{-1}\left[t_\sigma^{R(L)} c_{k\sigma}^R \pm t_\sigma^{L(R)} c_{k\sigma}^L\right]$ where $\mathcal{T}_\sigma = \sqrt{|t_\sigma^L|^2 + |t_\sigma^R|^2} = \sqrt{2}|t_\sigma|$), taking into account the approximations introduced in Sec. I. The superscript $e$ $(o)$ refers to the even (odd) combination of the electrodes' operators. This transformation decomposes the total Hamiltonian into two independent parts, among which only one, involving the 'even' operators of the electrodes, represents the quantum dot coupled now to a single reservoir of electrons. This is because the tunnelling Hamiltonian takes the form

$$\mathcal{H}_{\text{tun}} = \sum_{\mathbf{k}\sigma} \mathcal{T}_\sigma d_\sigma^\dagger c_{\mathbf{k}\sigma}^e + \text{H.c.}, \tag{S-19}$$

with the new effective tunneling matrix element $\mathcal{T}_\sigma$. This significantly simplifies numerical calculations by reducing effectively the coupling of the dot to a single conduction-electron channel, the regime in which high-spin molecular devices usually operate in experiments[32,33]. Moreover, the transformation shows that the spin dependence can be taken into account entirely *via* the matrix element $\mathcal{T}_\sigma$[23]. The conduction band is approximated as being flat in the interval $[-W, W]$, so that the DOS becomes $\nu_\sigma \equiv \nu = \frac{1}{2W}$.

The spectral function plays a key role: not only does it fully determine the linear-response (equilibrium) transport properties reported in the main article, but it also allows one to draw some qualitative predictions about the system's behaviour at finite bias voltage. In particular, the differential conductance of the system at low $T$ and low, finite bias voltage $V_b$ can be approximated by the symmetrised equilibrium spectral function,

$$\frac{dI}{dV_b} \propto A(\omega = V_b) \quad \text{with} \quad A(\omega) = \pi\Gamma \sum_\sigma (1 + \eta_\sigma p)\frac{a_\sigma(\omega) + a_\sigma(-\omega)}{2}. \tag{S-20}$$

Taking into account known limitations of this approximation[34-36], we thus use the spectral functions obtained by means of the DM-NRG approach to infer useful conclusions regarding spectroscopic signatures of spintronic anisotropy.

## III. SUPPORTING DISCUSSION AND FURTHER RESULTS

### A. Bias dependence of exchange fields

A key point of the analysis of the spectral signatures of the spintronic anisotropy in the following subsections is that its dependence on the level position $\varepsilon$ (controlled by the gate voltage $V_g$) is very different from the dependence on the bias voltage $V_b$. Here we first show that the bias dependence is in fact negligible for the regimes of interest. This can be done analytically within the RTD approach, illustrating, however, quite generic arguments.

As explained at the end of Sec. II A we can extend equation (2) and (3) to the case of a junction with two parallel ferromagnets at electrochemical potentials $\mu_r$ by simple substitutions and adding summations. The two functions $B_0$ and $B_2$ that make up both $B$ and $D$ [cf. equation (2) and (3)] then become

$$B_n = \frac{p\Gamma}{2\pi} \sum_{r=L,R} \mathcal{P} \int_{-W}^{W} d\omega \, \frac{f(\omega - \mu_r)}{\omega - \varepsilon - nU/2} = \frac{p\Gamma}{2\pi} \sum_{r=L,R} \ln \left| \frac{\varepsilon - \mu_r + nU/2}{W + \varepsilon + nU/2} \right|. \tag{S-21}$$

In the second step we approximated the Fermi function $f(\omega - \mu_r)$ by the step function $\theta(-\omega + \mu_r)$ since we focus on the regime of deep Coulomb blockade where thermal fluctuations in equation (S-21) become irrelevant. Assuming $|\varepsilon + U/2|, V_b \ll U \ll W$, the leading order terms in the expansion of the dipolar and quadrupolar exchange fields around the symmetry point in the variables $\varepsilon + U/2$ and $V_b$ are respectively given by:

$$B(\varepsilon, V_b) \approx B(\varepsilon) \left[ 1 + \left( \frac{V_b}{U} \right)^2 \right] \quad \text{and} \quad D(\varepsilon, V_b) \approx D^* \left[ 1 + \left( \frac{V_b}{U} \right)^2 \right], \tag{S-22}$$

where $B(\varepsilon) = -\frac{4}{\pi} p\Gamma \frac{\varepsilon + U/2}{U}$ and $D^* = -\frac{4}{\pi^2} \frac{(p\Gamma)^2}{U} \ln \frac{2W}{U}$ are the expressions (4) and (5) discussed in the main article, extended to the case of two electrodes. This indicates that there are three compelling reasons to neglect the bias dependence of the exchange fields. First of all, the relevant bias energy is the large energy scale $U$, i.e. the bias voltage corrections can generally be neglected as long as $V_b \ll U$. Second, the leading correction to *both* $B$ and $D$ is *quadratic* since the linear term vanishes by cancellation for symmetric bias (see below). There is no competition between a linear and quadratic dependence as discussed for the dependence on the level position $\varepsilon$. Third, the ratio of $B$ over $D$ is *constant* up to the second order in $V_b/U$, i.e. the *relative* importance of the dipolar and quadrupolar fields – of central interest in the main article – is thus affected only by terms of at least third order:

$$\frac{B(\varepsilon, V_b)}{D(\varepsilon, V_b)} \approx \frac{B(\varepsilon)}{D^*} + O\left( (V_b/U)^3 \right). \tag{S-23}$$

The above argument can be extended to include an experimentally relevant dependence of the level position $\varepsilon$ on both the gate and bias voltage: as long as the condition $\varepsilon(V_g, V_b) = -U/2$ is maintained, the effect of $\mu_L - \mu_R = V_b$ can be neglected. The non-equilibrium effect of $V_b$ is thus negligible, whereas the effect of $V_b$ on the level position can be compensated by the gate voltage. This corresponds to a simple 'skewing' of the quadrupole-dominated regions in the stability diagrams in Fig. S-11(a).

### B. Spectroscopic signature of spintronic anisotropy

A central question addressed by the paper is how far into the strong-$\Gamma$, low-$T$ regime the RTD prediction of a quadrupolar term in the effective Hamiltonian (1) of the spin-valve remains valid, even qualitatively. For this reason we applied the DM-NRG method to the same high-spin model. However, at present there is no direct way of calculating an effective subsystem Hamiltonian using this method. Therefore we must resort to carefully inferring the presence of spintronic dipolar and quadrupolar terms in equation (1) from their *effect* on quantities that are accessible with DM-NRG, such as the current and differential conductance, their spin polarization and local quantities such as the average spin and quadrupole moment.

The discussion of the question about how the *theoretical* DM-NRG results demonstrate that there is an effective non-zero quadrupolar term, naturally ties in with a second, central question of the main article: How can one *experimentally* prove – based on the same physical quantities, but now their measured values – that, e.g., a splitting of a feature in the differential conductance $dI/dV_b$, is in fact of spintronic origin and not intrinsic (i.e. due to spin-orbit coupling)? In the following two subsections we will answer both these questions by considering the features in the DM-NRG local density of states (S-20), a first approximation to the $dI/dV_b$ spectrum. We proceed in two steps: first, we review and discuss how to determine the sign of the underlying $D$ parameter, and second, we discuss which dependencies on experimental parameters enable one to identify this parameter as spintronic, i.e. induced by transport, rather than intrinsic.

### 1. Identification of quadrupolar term from transport spectra

Transport spectroscopy has been successful in accessing spin-1 states as excitations of various types in quantum dot systems[4,5,8,9,32,33,37–40]. To this end one plots the differential conductance $dI/dV_b$ as a function of the applied bias voltage and the gate voltage (level position). We first review how the inelastic cotunneling (or IETS) $dI/dV_b$ can reveal a uniaxial anisotropy of the form (1) for a quantum dot with an $S = 1$ ground state, regardless of the origin of this anisotropy, either *intrinsic* or *spintronic*. In general, at zero magnetic field the presence of fixed magnetic anisotropy ($D \neq 0$) manifests as a splitting of the Kondo resonance[41,42] into peaks at the renormalized, finite excitation energies with logarithmic enhancements[40,43,44]. For larger values of $D$ the remnants of the split Kondo peak evolve into inelastic cotunneling steps with slight non-equilibrium overshooting at the step-edge[35,45–48]. It was noted experimentally in Ref. [32] that just from these features the *sign* of the underlying $D$ parameter cannot be determined easily and that a tunable, external magnetic field $B$ becomes crucial.

To illustrate how a magnetic field can be employed to distinguish between the so-called *'easy axis'* ($D < 0$) and *'easy plane'* ($D > 0$) magnetic anisotropy, we now first ignore the complications of the Kondo effect (included in our DM-NRG calculations). We discuss the case of a spin $S = 1$ with *intrinsic* uniaxial anisotropy [described by equation (1)] and weakly **Heisenberg**-coupled to tunneling electrons, see Fig. S-1. The non-linear conductance is obtained using the leading order perturbation theory (Fermi golden rule) and a master equation, for details see Refs. [46–48]. In Fig. S-1(a) we show that in the absence of an external magnetic field the inelastic cotunneling spectra are qualitatively the same for both signs of $D$. Then, an external magnetic field engenders qualitative differences, see Fig. S-1(b), due to the Zeeman splitting of the ground and the excited state, respectively, as Fig. S-1(c) demonstrates. Specific to the case $D < 0$ is that the increase of the external field $B_z$ along the spin's easy axis is accompanied by an increase of the gap *without splitting* the excitation. To identify unambiguously the sign of the magnetic anisotropy constant $D$, one thus needs to investigate the evolution of the spectral signatures as a function of the magnetic field $B_z$. Experimentally, such a behaviour was indeed observed for magnetic adatoms[49–51] as well as single-molecule magnets[35,52,53] in a real, external magnetic field[54].

In the case of interest in the manuscript, both the anisotropy and magnetic field are of *spintronic origin*. As a result, the exchange field $B$, equation (2), can be tuned approximately linearly with the gate voltage around $\varepsilon = -U/2$ and is always collinearly aligned with the anisotropy axis (both the direction of $B$ and the axes of $D$ are set by parallel axes of the magnetic moments of electrodes). The theoretical DM-NRG gap $\Delta E$ shown in Fig. 3(b) of the main article and in Fig. S-2(b), displays a linear increase without any additional splitting, see the caption of Fig. S-2(b) for details on the determination of $\Delta E$. This confirms that we have an *'easy-axis'* anisotropy ($D < 0$) underlying the DM-NRG calculations. Clearly, this identification can also be made in an experimental situation if the differential conductance resembles Fig. 3(b). Particularly interesting from the experimental point of view is that the direction of the easy axis of the spintronic anisotropy is *known* from the ferromagnets' polarizations and is controllable by a strong, externally applied (real) magnetic field that reorients the ferromagnets' polarization and is then switched off. This stands in contrast to intrinsic spin-orbit induced anisotropy, which relies on the orientation of single-molecule magnet on a surface or the detailed atomic structure surrounding an adatom.

In the theoretical DM-NRG calculations, we can furthermore independently infer the sign of $D$ from the sign of the positive value of $\langle \hat{Q}_{zz} \rangle$ plotted in Fig. 4(b) of the main article: an elementary

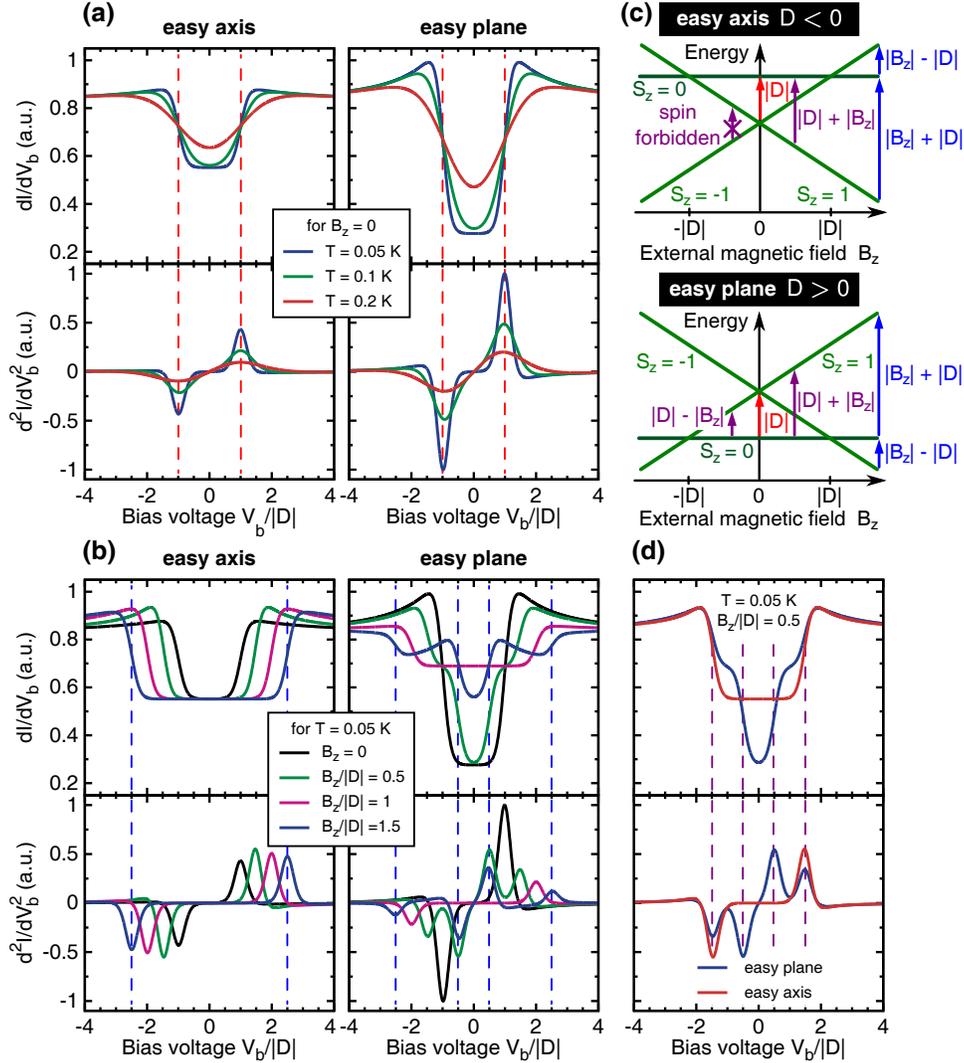

FIG. S-1. **Spectral signatures of the anisotropy sign:** Inelastic cotunneling spectra, i.e. differential conductance $dI/dV_b$ vs. bias voltage $V_b$, for two different signs of the *intrinsic* magnetic anisotropy constant corresponding to so-called '*easy axis*' ($D < 0$) and '*easy plane*' ($D > 0$) magnetic anisotropy. For illustration we use $|D| = 50$ μeV . The inflection points in $dI/dV_b$ associated with magnetic resonances are easily identified using the shown second derivatives of the current $I$. (a) Zero magnetic-field temperature dependence. (b) Low-temperature magnetic-field dependence. (c) Dependence of the spin's energy spectrum on an *external* magnetic field $B_z$ along the easy axis with color arrows denoting exchange-tunneling induced transitions and associated transition energies. (d) Direct comparison between the cases of '*easy axis*' and '*easy plane*' magnetic anisotropy. Note that in (a)-(b) and (d) the dashed lines serve as a guide for eyes. The color of each guiding line corresponds to the color of the arrow indicating the transition in sketch (c).

calculation shows that it is a direct consequence of having a negative anisotropy term $D < 0$ in the effective Hamiltonian.

Finally, it should be noted that transverse magnetic anisotropy (not present here) can be distinguished by known different additional signatures: an exotic Kondo effect[56,57] at *zero* bias voltage and field and weak non-equilibrium effects within the gap, see supporting information of Ref. [35].

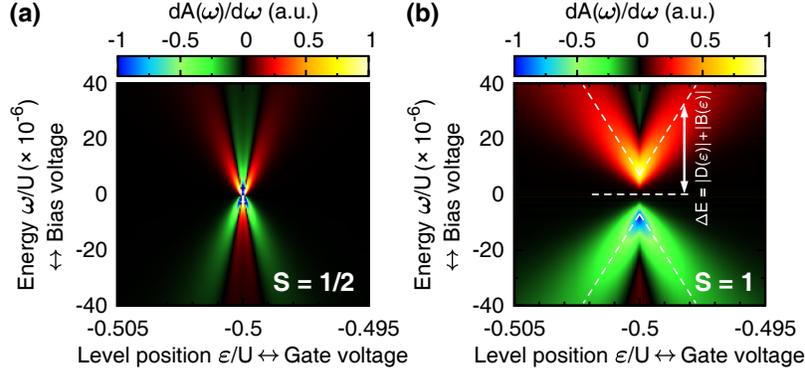

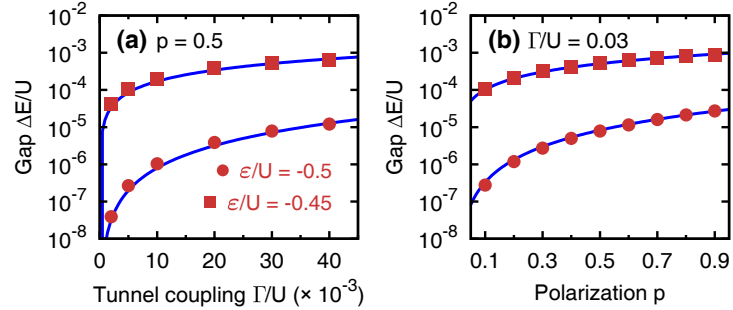

### 2. Spintronic origin of quadrupolar field

Once a quadrupolar gap with a negative signature of $D$ has been identified, one may conclude its spintronic origin by checking the three following properties, two of which are plotted in Fig. S-3.

1. **$\Gamma$-dependence:** The scaling of the energy axis in Fig. 3(c) [compare with Fig. S-3(a)] reveals that the DM-NRG gap indeed scales with $\Gamma$ as predicted by equations (4)-(5). Experimentally, this should appear as a clear dependence on the electrode distance in scanning-probe experiments, as well as in two- and three-terminal mechanically controllable molecular junctions[58–60].

2. **$p$-dependence:** According to Fig. 3(d) [compare with Fig. S-3(b)] the DM-NRG gap scales as predicted with the electrodes' spin polarization $p$. Experimentally, this may be tested conveniently in scanning-probe experiments where greater material/technique flexibility of the magnetic tip's preparation is available at the moment[61].

3. **$\varepsilon$-dependence:** In general, the gate voltage dependence of the exchange fields are their most characteristic feature. For the dipolar exchange field $B$ this dependence was clearly mapped out experimentally[62]. The gate voltage-dependence of $D$, however, becomes pronounced in the regime where $B$ dominates over $D$, as shown in Fig. 2(a) of the main article. An external magnetic field $B_{\text{ext}}$ is required to access the values of $D(\varepsilon)$ for $\varepsilon \neq -U/2$, see discussion of equation (S-31) below.

### C. Vanishing of dipolar exchange field and average spin at the symmetry point

Next, we elaborate on the vanishing of the dipolar exchange field at the symmetry point. The dipolar exchange field was first obtained for the non-equilibrium spin-dynamics[21] of spin-valves and subsequently also in the case of the Kondo effect[63,64]. One of the key features of this exchange field is that it exhibits quantum interference between electron and hole tunneling processes, which results in a vanishing of the field at the symmetry point $\varepsilon = -U/2$ at zero bias. It relies on the electron-hole symmetry properties of the composite system of quantum dot plus ferromagnets[23]. In general, depending on the large-scale energy-dependence of the density of states of the ferromagnets, this compensation point can shift. Though these considerations concerned low-spin $S = 1/2$ quantum-dot spin valves, they in principle apply in order $\Gamma$ also to a $S = 1$ system[22], cf. equation (2).

Earlier NRG calculations[23,63,65] have already demonstrated that for $S = 1/2$ higher-order $\Gamma$ corrections indeed preserve the above property of $B$. Thus, the cancellation of the dipolar exchange field at the symmetry point is a generic feature of interacting quantum dot spin-valves if the effects of spin polarization dominate over those due to the magnetization of ferromagnets, see Ref. [66]. Our DM-NRG calculations indicate that this also remains true for $S = 1$, i.e. we find that for $\varepsilon = -U/2$ the average spin is exactly zero:

$$\langle \hat{S}_z \rangle = 0 \quad \text{for} \quad \varepsilon = -U/2, \tag{S-24}$$

whereas $\langle \hat{S}_x \rangle = \langle \hat{S}_y \rangle = 0$ follows from symmetry with respect to spin-rotations about the common spin-polarization axis of the ferromagnets (i.e. the $z$-axis). Note that if an effective dipolar field $B$ was present at the symmetry point, it would certainly split the Kondo resonance in Fig. S-4(b). Moreover, an extensive parameter scan confirms that at $\varepsilon = -U/2$ this does not take place for any set of parameters $\Gamma$, $U$, $p$. This holds despite the added complexity due to the quadrupolar exchange field.

### D. Competition between spintronic anisotropy and Kondo effect

As it was mentioned in the main article, the Kondo effect in Fig. 3(b) reinstates the characteristics similar to those of a $S = 1/2$ system shown in Fig. 3(a), but only for large tunnel coupling $\Gamma$. Here we discuss this competition with the superparamagnetism in detail, controlled by the experimental parameters $T$ and $\Gamma$.

Despite the zero average spin at the symmetry point $\varepsilon = -U/2$ [cf. equation (S-24)], a quantum dot with a spin $S > 1/2$ can still be found to have a non-paramagnetic state. This non-trivial superparamagnetic spin-state is characterized by

$$\langle \hat{Q}_{zz} \rangle > 0 \quad \text{for} \quad \varepsilon = -U/2. \tag{S-25}$$

In Fig. S-4 we discuss the appearance of this superparamagnetism in the energy (bias) dependence as the tunnel coupling is varied. Then, in Fig. S-5 we show a full $(T, \Gamma)$-diagram. As mentioned in the main article (but not shown there), the spintronic anisotropy can also be determined from its signatures in the temperature dependence of various measurable quantities. This is illustrated in Fig. S-6. In all these figures three regimes can be distinguished, denoted by (i)-(iii) as in Fig. 4 and its caption in the main article.

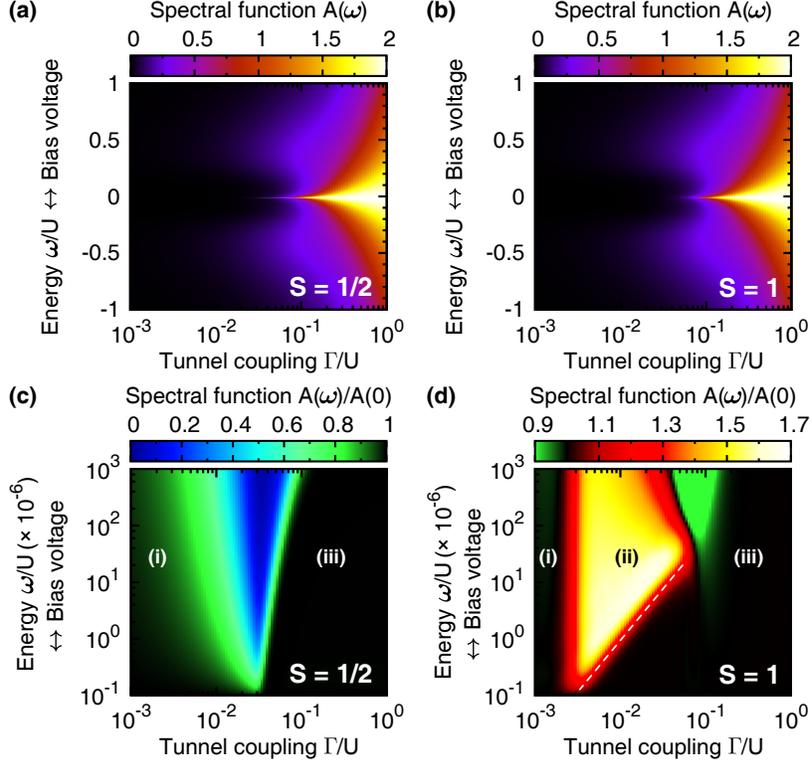

FIG. S-4. **Comparison of spectral functions for a low- and high-spin quantum-dot spin-valve:** Spectral function $A(\omega)$ at the symmetry point $\varepsilon = -U/2$ for (a) spin $S = 1/2$ [$K = 0$ in equation (S-1)] and (b) $S = 1$ [$K < 0$ in equation (S-1)]. In both cases sufficiently large tunnel coupling $\Gamma$ results in a zero-bias Kondo peak between the Hubbard peaks around $\omega \approx \pm U$. However, only for $S = 1/2$ the Kondo peak persists for arbitrarily small $\Gamma$, albeit with an exponentially vanishing width. For $S = 1$, on the other hand, this does not hold as the bottom panel demonstrates, where we plot relevant zoom-ins of (a) and (b), using a logarithmic energy scale. We normalize $A(\omega)$ to $A(0)$, the spectral function at the smallest energy resolvable by the DM-NRG (i.e. at $\omega \sim T$). Note that $A(0)$ depends on $\Gamma$. One now clearly distinguishes only two regimes for $S = 1/2$ in (c) but three regimes for $S = 1$ in (d). (i) *Paramagnetic regime*: Initially, for small $\Gamma$ we have $T \gtrsim |D^*(\Gamma)|$ and $T_K(\Gamma)$, so that $A(\omega)$ is a featureless elastic tunneling curve. (ii) *Superparamagnetic regime*: In (d) for intermediate $\Gamma$-couplings a quadrupolar gap arises, scaling as $\propto \Gamma^2$, for which the spectral function at finite energy $\omega \gtrsim |D^*(\Gamma)|$ is large relative to $A(0)$. This regime is missing in (c). (iii) *Kondo screened regime*: Ultimately, for large $\Gamma$ there is a Kondo peak at low energy and the situation reverses, i.e. $A(\omega) < A(0)$, see plots (a)-(b). As in Fig. S-2 the dashed line in (d) indicates the position of the inflection point of $A(\omega)$, being the measure of the quadrupolar gap $\Delta E = |D^*|$. To illustrate all possible regimes in a single calculation we assumed $T/W = 10^{-9}$.

(i) **Paramagnetic regime:** Initially, for small tunnel coupling $\Gamma$ all energy scales fall below temperature $T$,

$$T \gtrsim |D^*(\Gamma)| \text{ and } T_K(\Gamma, D^*(\Gamma)), \tag{S-26}$$

with $T_K(\Gamma, D^*(\Gamma))$ denoting the Kondo temperature (see below), and the spin behaves as an (isotropic) paramagnet, i.e. $\langle \hat{S}_z \rangle = 0$ and $\langle \hat{Q}_{zz} \rangle = 0$.

(ii) **Superparamagnetic regime:** For a given, finite temperature $T$ the superparamagnetic regime, where $\langle \hat{Q}_{zz} \rangle > 0$, is reached upon increasing the tunnel coupling $\Gamma$ from zero when

$$|D^*(\Gamma)| \gtrsim T \text{ and } T_K(\Gamma, D^*(\Gamma)), \tag{S-27}$$

i.e. the quadrupolar gap exceeds the thermal energy scale $T$. The condition $|D^*(\Gamma)| = T$ defines the sloped boundary between (i) and (ii) in Fig. S-5. The DM-NRG calculations

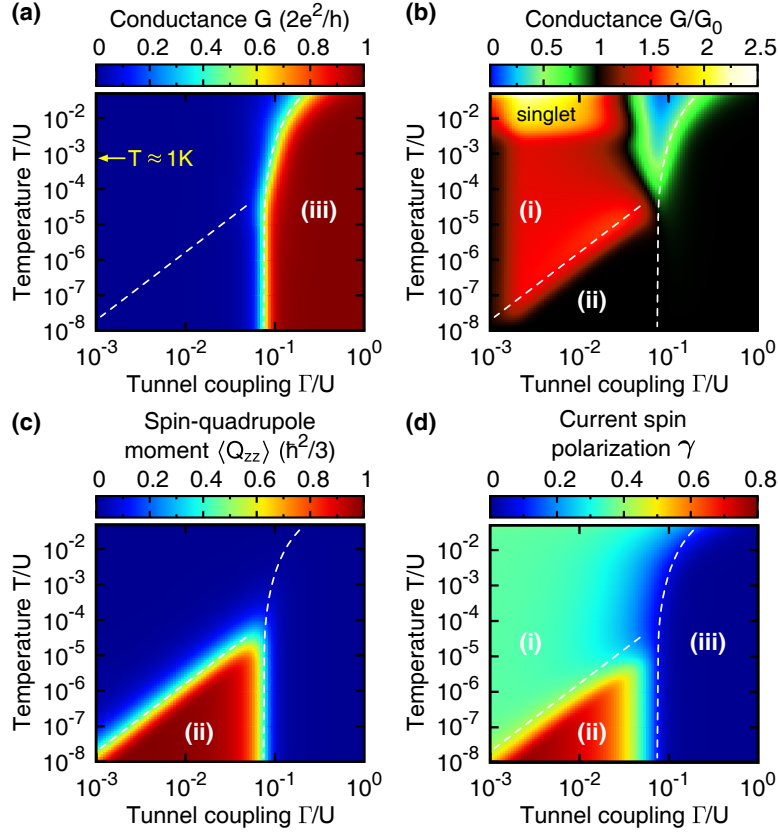

FIG. S-5. **Competition between spintronic anisotropy and Kondo effect:** The linear transport through an $S = 1$ quantum dot spin-valve tuned to the symmetry point $\varepsilon = -U/2$ is presented for $p = 0.5$. Various physical observables that distinguish the different magnetic regimes, cf. Fig. S-4, are shown plotted as functions of $(\Gamma, T)$. The dashed lines are merely a guide for eyes, and have exactly the same relative position on all the plots to enable direct comparison of plots (a)-(d). (a) The linear conductance $G = G_\uparrow + G_\downarrow$ shows a transition to the Kondo regime (iii). For high temperatures, $T \gtrsim |D^*(\Gamma)|$, the boundary depends on $T$. Below this threshold, however, the boundary becomes vertical, i.e. $T$-independent and its position is determined by the condition $|D^*(\Gamma)| \sim T_K(\Gamma, D^*(\Gamma))$. (b) The linear conductance $G$ – now normalized to its low-temperature value $G_0 \equiv G(T/U = 10^{-8})$ – additionally reveals the opening-up of the quadrupolar gap, which shows up as the border between regimes (i) and (ii). (c) The quadrupole moment $\langle \hat{Q}_{zz} \rangle$ (normalized to its maximal value) also shows a transition to the superparamagnetic regime (ii) along the sloped line where $T \sim |D^*(\Gamma)|$. Note that above the 'cusp' of regime (ii) in this figure the transition to the Kondo regime in plot (a) becomes $T$-dependent. (d) The current spin polarization $\gamma = (G_\uparrow - G_\downarrow)/(G_\uparrow + G_\downarrow)$ allows one to most clearly distinguish the paramagnetic regime (i). As noted in the caption of Fig. 4 of the article, lower temperatures are required to obtain enhanced spin filtering as compared to having a finite quadrupole moment: compare the boundaries of regime (ii) in (d) and (c), respectively. See also Sec. III F and the caption of Fig. S-10. The comparison with corresponding values for the $S = 1/2$ case shown in Fig. 4(c) of the main article demonstrates that the system behaves as a free spin.

confirm that this line shifts up uniformly when increasing the spin polarization of electrodes $p$ [see the center panel of Fig. S-6], while it shifts down upon raising the Coulomb energy $U$ [see the right panel of Fig. S-6], as predicted by equations (S-29) and (S-30) below.

(iii) **Kondo screened regime:** On the other hand, Kondo exchange processes tend to destroy the superparamagnetic state. For the situation of interest here the singlet state at energy $\sim |K|$ plays no role. We note that the triplet-singlet excitation can be identified as a bright spot in the left-hand side corner of Fig. S-5(b). As explained in the caption of Fig. S-6, the Kondo scale $T_K$ then depends not only on $\Gamma$, $U$ (as usual[68]) and the spin polarization $p$[63,69],

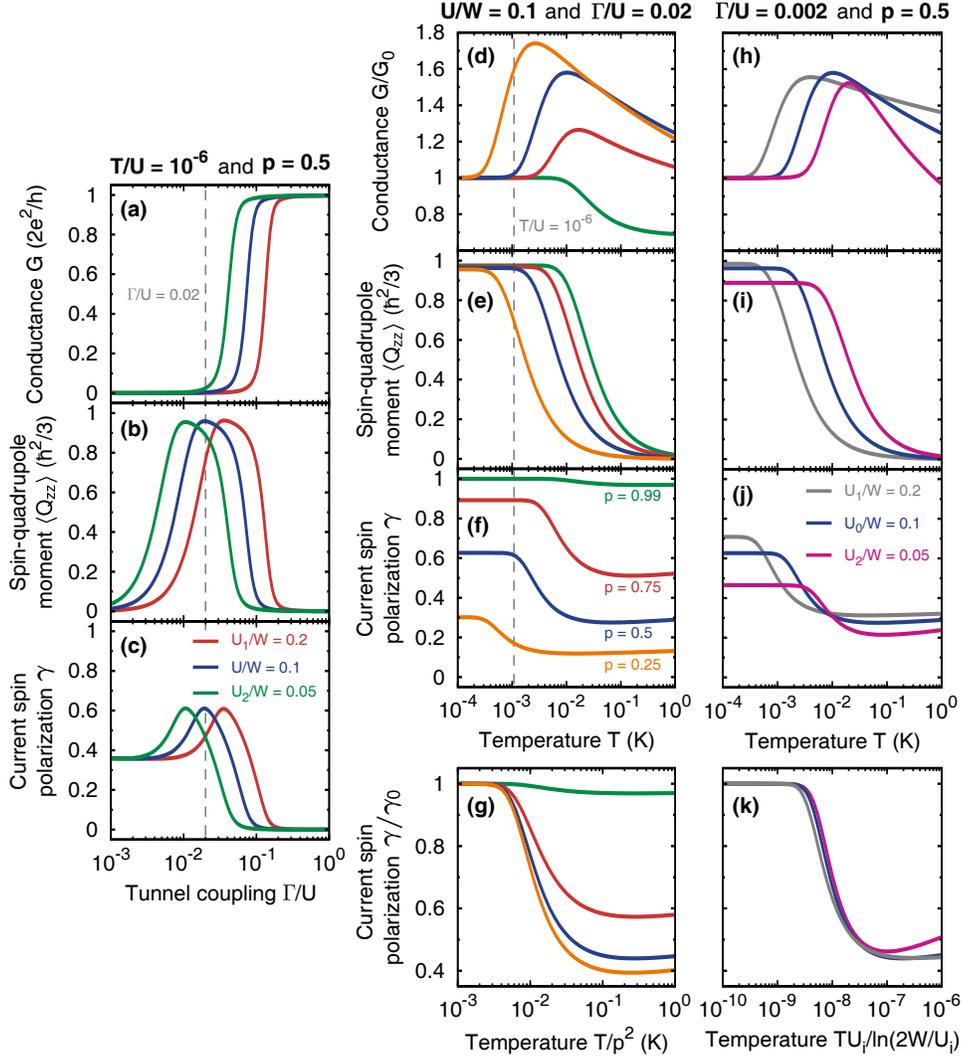

FIG. S-6. **Influence of Coulomb interaction and spin polarization on the spintronic super-paramagnetism:** *Left panel*: Dependence of relevant horizontal cross-sections of plots in Fig. S-5 for the indicated value of temperature $T$ on the Coulomb interaction $U$. It shows that the increase of $U$ results in the shift of the onset of the plateaus in (b)-(c) towards larger values of $\Gamma$, which corresponds to the decrease of $D$. On the other hand, the drop of the plateaus and concomitant saturation of the conductance in (a) prove that the Kondo temperature is also a function of $D$, i.e. $T_{\mathrm{K}}\big(\Gamma, U, D^\star(\Gamma, U)\big)$. Otherwise the onset of the Kondo effect, determined simply by $|D^\star(\Gamma, U)| = T_{\mathrm{K}}(\Gamma, U)$, would occur at a fixed ratio of $\Gamma/U$. Recall that for the symmetric single-impurity Anderson model[67] with $S = 1/2$: $T_{\mathrm{K}}^{(1/2)}(\Gamma, U) = \frac{1}{2}\sqrt{\Gamma U}\exp\big[-\pi U/(4\Gamma)\big]$. *Center panel*: (d)-(g) Temperature dependence of the same quantities as above, i.e. vertical cuts through the corresponding plots in Fig. S-5, shown for different values of spin polarization $p$. Conductance and current spin polarization data are normalized to their values at temperature $T = 10^{-5}$ K. In the conductance a low-temperature (Kondo enhanced) peak appears at $T = |D^\star|$, which for very high spin polarization $p$ turns into a step due to the loss of Kondo correlations. Plot (e) shows that this feature indeed agrees with the onset of the quadrupole moment induced by the **quadrupolar field**. The spin-filtering enhancement by the quadrupolar field, discussed in Fig. 4 of the main article, also shows up in the temperature dependence of the current spin-polarization $\gamma$ in panel (f). Finally, plot (g) confirms the scaling with the polarization $p$ given by equation (5) of the main article, but now for the $T$-dependent signatures. *Right panel*: Analogous to the center panel, except that the dependence on the Coulomb interaction $U$ is now shown.

but also on the anisotropy splitting $D^*(\Gamma)$, which in turn depends on $\Gamma$. In consequence, as a function of $\Gamma$, the Kondo regime is thus reached only when $\Gamma$ satisfies the equation:

$$T_{\mathrm{K}}\big(\Gamma, D^*(\Gamma)\big) \gtrsim |D^*(\Gamma)|. \tag{S-28}$$

The distinguishing feature of this condition is that, as long as equation (S-27) holds, it is independent of temperature $T$, which corresponds to the vertical (ii)-(iii) boundary in Fig. S-5. The DM-NRG calculations furthermore confirm that the $\Gamma$-position of this vertical line shifts to the right side when increasing spin polarization $p$ as predicted by equations (S-29) and (S-30) below, see Fig. S-10: in those cases the Kondo exchange processes are suppressed (as is known from earlier studies[63]).

## E. Estimation of experimentally achievable spintronic anisotropy

In the main article we restricted the parameters employed for calculating Figs. 2-4 to enable comparison of results obtained by the two complementary methods that we use, the RTD technique (Sec. II A) and DM-NRG approach (Sec. II B). Here we present estimations of the experimentally achievable energy scale for $D^*$ corresponding to the symmetry point $\varepsilon = -U/2$. We start from the relation (5) extended to the case of two electrodes, i.e.

$$|D^*_{\mathrm{RTD}}| \approx \eta_{\mathrm{RTD}} \times p^2 \frac{\Gamma}{U}\Gamma \quad \text{with} \quad \eta_{\mathrm{RTD}} \approx 0.41 \times \ln\left(\frac{2W}{U}\right). \tag{S-29}$$

The logarithmic factor is only valid for $U \ll W$ and provides an enhancement, i.e. $\ln(2W/U) > 1$. Using the DM-NRG we can calculate the gap $|D^*|$ for arbitrary $W/U$. Since the dependence on $W$ is very weak (logarithmic) we now discuss the parameter dependence for the fixed ratio $W/U = 10$ used in the main article. Using the DM-NRG we find numerically that

$$|D^*_{\mathrm{NRG}}| \approx \eta_{\mathrm{NRG}} \times p^2 \frac{\Gamma}{U}\Gamma \quad \text{with} \quad \eta_{\mathrm{NRG}} \approx 0.037. \tag{S-30}$$

The coefficient $\eta_{\mathrm{NRG}}$ has been obtained by fitting the above equation (with error less than 1%) to the data points (bullets) in S-3(b), and for comparison $\eta_{\mathrm{RTD}} \approx 1.21$. The difference in the coefficients is expected for various reasons: (i) Equation (S-29) neglects a correction of the first order relative to $\ln(2W/U)$ [see equation (S-11)] which tends to decrease $|D^*|$; (ii) In the RTD calculation we excluded the singlet excited state, which further reduces $|D^*|$. Equation (S-29) thus presents an upper bound in the limit of weak tunnel-coupling. The $|K|$ dependence is explored in Fig. S-7, recovering $D^* = 0$ for $K = 0$ as expected ($S = 1/2$ case). (iii) Estimation of $|D^*|$ from the DM-NRG results is based on the low-energy inflection point of the spectral function, which gives a lower bound to $|D^*|$ [see discussion of equation (S-31) below].

The achievable magnitude of the quadrupolar gap $|D^*|$ then relies essentially on two factors whose experimental values we discuss now:

1. **Junction parameters $\Gamma$ and $U$:** The maximal achievable tunnel coupling $\Gamma$ is probably the most tunable parameter of all and as a rule it increases as the dimension of a quantum dot device is reduced. Notably, the ratio $U/\Gamma$, larger than 2 in typical experiments, is quite similar for a wide range of quantum dots with drastically different absolute values of $\Gamma$. For instance, in semiconductor quantum dots[5,39,70–75] $\Gamma \lesssim 0.5$ meV and $U$ hardly ever exceeds a few meV. In carbon-nanotube quantum dots[62,66,76–78] $\Gamma$ and $U$ can vary in a significantly broader range with $U$ even reaching tens of meV[79,80] (typically $U \approx 5-20$ meV/L[$\mu$m][81,82] depending on the length $L$ of the nanotube captured between the contacts). In molecules and adatoms one expects that $U$ achieves values on the eV scale, and this can indeed be observed[9,83–85]. However, the proximity to metal electrode surfaces not only enhances $\Gamma$, but also leads to significant diminishing of molecular energy scales due to image charge and polarization effects[86]. For this reason, molecular quantum dots can also show rather low $U$ values depending on junction details[87], e.g., $U \sim 30-40$ meV and $\Gamma \sim 6-35$ meV (depending on the charging state) in Ref. [88]. A strong dependence of the transport gap (the effective HOMO-LUMO gap to which $U$ contributes) on the electrode spacing has been demonstrated in three terminal mechanically-controlled break junctions[58–60]. Thus, large values of tunnel coupling $\Gamma$ with restricted ratios of $\Gamma/U$ can be achieved in nanoscale quantum dots and make them attractive for generating spintronic anisotropy.

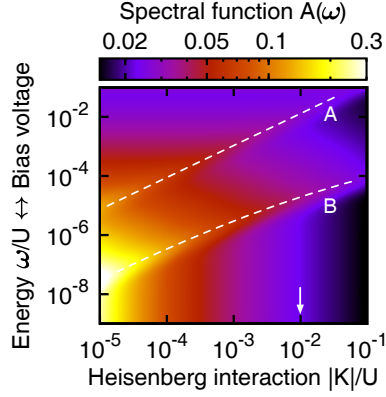

FIG. S-7. **Influence of the Heisenberg interaction on the spintronic anisotropy:** Spectral function $A(\omega)$ at the symmetry point $\varepsilon = -U/2$ shown as a function of the strength of the ferromagnetic Heisenberg interaction $|K|$, with other parameters the same as in Fig. 3(b) of the main article, the white arrow indicating the value of $|K|$ used there. The dashed lines mark the position of peaks associated with (A) the triplet-singlet excitation at energy $|K|$ and (B) the quadrupolar gap $|D^*|$. Our spintronic mechanism allows for an anisotropy parameter whose magnitude $|D^*|$ increases with the triplet-singlet excitation gap $|K|$: the better defined the high spin is, the more anisotropic it becomes. By adjusting the tunnel coupling ($\Gamma$) and polarization ($p$) the anisotropy $|D^*|$ can be made a sizeable fraction of $|K|$. This adjustment involves a non-trivial dependence of the interplay with the Kondo effect, see in Sec. III D. We note that for any anisotropic spin system, either real single-molecule magnets/adatoms, or spintronic single-molecule magnets introduced in our article, it is of no use to have an anisotropy value $|D^*|$ close to or larger than $|K|$: in this case $|K|$ becomes the limiting factor for applications (spin-flips destroy the large spin), rather than $|D^*|$ (large spin is reoriented). Finally, we also note that increasing the gap $|D^*|$ to the excited spin state – *even* when this state is already out of reach – is advantageous since it enhances the spintronic anisotropy further.

2. **Electrode spin-polarization** $p$: The maximal value of the spin-polarization factor $p^2$ in equations (S-29) and (S-30) presents the limiting factor for achieving large spintronic anisotropy. Naively, one may expect it to be set by the type of material used for fabricating electrodes. In general, for metallic ferromagnetic metals[89,90] one can find $p \sim 0.4 - 0.6$, while for half-metallic ferromagnets (e.g., some Heusler compounds, zinc-blend structure materials, and some magnetic oxides) at low temperature $p$ can be even larger, reaching in principle the limit of $p \sim 1$, see Chap. 5 of Ref. [91]. However, low-temperature experiments on junctions involving magnetic electrodes turned out to be challenging, reporting lower values of $p$: in experiments with carbon nanotubes $p \approx 0.1$ for contacts made from PdIn alloy[66], $p \approx 0.2$ for PdNi alloy[92] or pure Ni[62,93], and $p \approx 0.3$ for Co[94]. In addition to material properties, in junctions created with a scanning probe the spin polarization of the tip can also be modified by different methods of preparation[61]. For instance, for tips from bulk ferromagnetic metal, such as Fe, a spin polarization $p \approx 0.4 - 0.5$ has been observed (and half of that value for Cr). On the other hand, spin polarizations of the order of 0.3 for a single magnetic apex atom (Fe or Mn) and 0.4 for multiple atoms attached to the tip apex (in external magnetic field) have been found hitherto[51]. This method has been suggested[95] to be able to produce $p \approx 0.8$. Clearly, here there is room for improvement and promising routes have been suggested.

Next, in order to estimate the maximal achievable magnitude of $|D^*|$ we need to investigate the non-trivial dependence of the spintronic anisotropy on tunnel coupling $\Gamma$ and the Coulomb interaction $U$. Fig. S-8 shows that due to the competition of the Kondo effect and the spintronic anisotropy, discussed in Sec. III D, the perturbative enhancement of $|D^*|$ with increasing $\Gamma$ and decreasing $U$, as predicted by equation (S-29), is stopped as soon as the Kondo effect is restored. This leads to a saturation of $|D^*|$ at the following approximate value, read off from Fig. S-8(b)-(c),

$$|D^*| = 0.02 - 0.04 \text{ meV} \quad \text{for} \quad K = -1 \text{ meV}, \ W = 1 \text{ eV}, \text{ and } p = 0.9, \quad \text{(S-31)}$$

which cannot be overcome by further tuning of $\Gamma$ and $U$. Note that for higher (lower) values of $p$ the Kondo effect is additionally suppressed (enhanced) and the saturation value $D^*$ of is higher (lower).

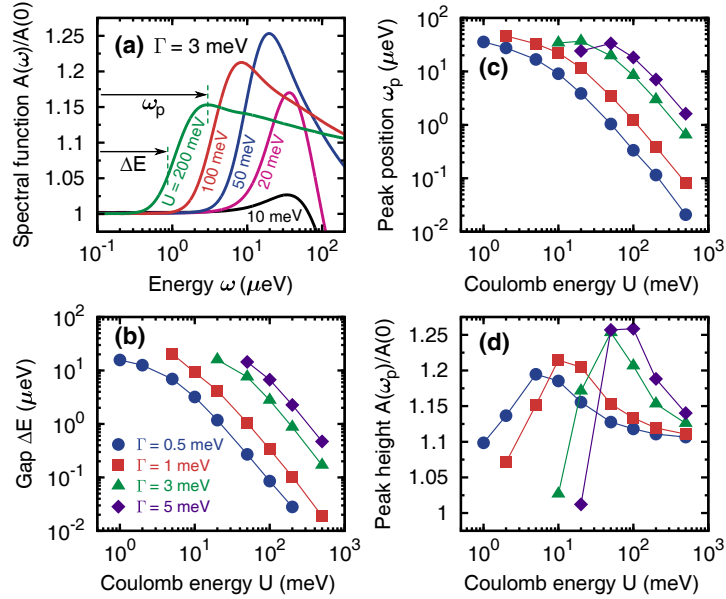

FIG. S-8. **Estimation of experimentally achievable spintronic anisotropy:** Dependence of the quadrupolar gap $\Delta E = |D^*|$ on the Coulomb interaction $U$ for several values of tunnel coupling $\Gamma$ and large spin polarization $p = 0.9$. The gap $\Delta E$ is extracted from the DM-NRG spectral function $A(\omega)$ in the same way as in Fig. S-3. (a) An example of the evolution of the spectral function with $U$: the quadrupolar feature in the inelastic tunneling can take the form of a stepped curve, a Kondo-enhanced peak or a combination of both. Note that the meaning of $A(0)$ is the same as in Fig. S-4. (b) The gap $\Delta E$ extracted from the low-energy inflection point. (c) Peak position $\omega_p$ of the quadrupolar spectral feature. (d) Peak height normalized to the zero-energy value $A(0)$ (cf. caption Fig. S-4). The inflection point in (b) provides the safest and most robust measure of the quadrupolar gap, since it tends to underestimate the magnitude $|D^*|$ and it is always present, in contrast to the peak, which may vanish, especially for small $U$, see plot (d). Since in many cases the step and peak width are comparable with the size of the gap $|D^*|$, there is a substantial difference in these two measures of the quadrupolar splitting.

Increasing (decreasing) $|K|$ also enhances (suppresses) $|D^*|$. The value (S-31) is promising in view of the values obtained for state-of-the-art single-molecule magnets used in transport experiments, e.g., $|D| \approx 0.056$ meV for a Fe$_4$ molecule embedded in a three-terminal device geometry[35,55] or $|D| \approx 0.06$ meV for a Mn$_{12}$ molecule grafted on an insulating BN monolayer on Rh and studied by means of scanning tunneling spectroscopy[53]. Such experiments are challenging since molecular magnets often lose their magnetic anisotropy when deposited on a metal surface as required for device applications[53,96–98]. In our spintronic approach the anisotropy is generated (rather than lost) when a high spin system is incorporated into a spin-valve device structure. Individual magnetic adatoms, which have been studied on various substrates[99], display a much wider range of the $D$ parameter values. In particular, their uniaxial anisotropy constant can be as small as for single-molecule magnets (e.g., $|D| \approx 0.04$ meV for a Mn adatom deposited on an insulating Cu$_2$N layer[50,100]), but it can also reach significantly larger values of the order of few meV (e.g., $|D| \approx 1.55$ meV for a Fe adatom[100] or $|D| \approx 2.75$ meV for a Co adatom[49] – both deposited on an insulating Cu$_2$N layer, and the latter one exhibiting the '*easy-plane*' type anisotropy). The above conservative estimate for the spintronic quadrupolar gap corresponds to an anomalous inelastic cotunneling gap at the symmetry point of $2|D^*| = 80$ $\mu$eV, which through $k_B T \sim 80/4$ $\mu$eV translates to a temperature $T \sim 230$ mK. For this temperature the gap should be clearly visible. In high-resolution experiments (using, e.g., lock-in techniques) we expect weaker signatures to appear already at higher temperatures. Note that in our estimate we accounted for a $4k_B T$ broadening of the Fermi-Dirac distribution of electrons, as well as for the factor 2 in front of $|D^*|$, the relevant splitting being that between the spectral function peaks at $V_b = \omega = \pm|D^*|$, as indicated in Fig. 3(b) of the main article.

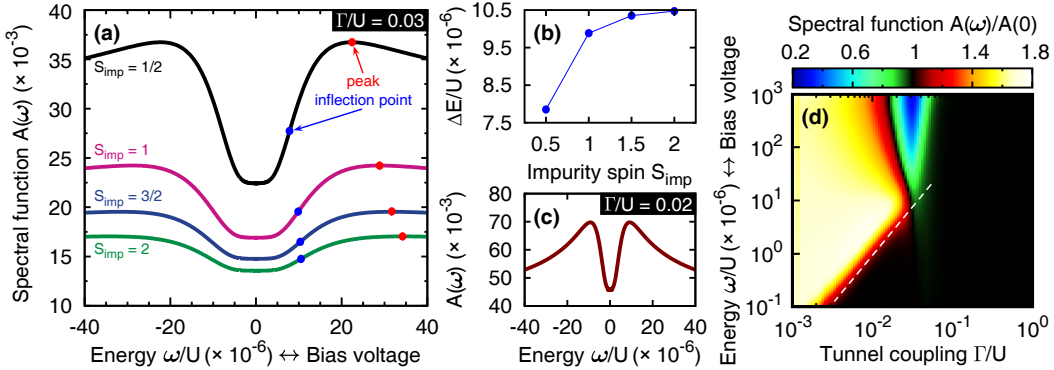

FIG. S-9. **Generic nature of the model: higher spin values and charge fluctuations of the impurity:** (a) Dependence of the spectral function $A(\omega)$ at the symmetry point $\varepsilon = -U/2$ on energy $\omega$ for models with different values of the impurity spin $S_{imp}$. Coloured dots mark the position of inflection points (blue dots) and peaks (red dots) of $A(\omega)$ [which is a symmetric function of $\omega$]. As previously, we use the inflection points as a conservative estimate of the quadrupolar gap $\Delta E = |D^*|$. Note that the black (top) curve in (a) is a cross-section of Fig. 3(b) of the main article. (b) Dependence of the inflection point energy position on the impurity spin $S_{imp}$. (c) Same as (a) but for the model of a two-level quantum dot with two orbitals equally coupled to a single ferromagnetic electrode, assuming for simplicity equal inter- and intra-dot charging energies equal to a large value of $U$. (d) The normalized spectral function $A(\omega)/A(0)$ for the two-level quantum dot shows the same *quantitative* behaviour as Fig. S-4(d) for an immobile impurity $S_{imp} = 1/2$: we superimposed the white dashed line indicating the inflection point of the *latter* figure. The only difference arises in the tunnel-coupling regime corresponding to a small portion of the dashed line extending beyond the red boundary of the yellow region. In order to facilitate the comparison between different cases the same parameters as in Fig. 3(b) of the main article were used for (a)-(b), while for (c) only a smaller value of the tunnel coupling $\Gamma$ had to be employed to avoid the mentioned earlier onset of the Kondo effect.

In order to complete the discussion of achievable magnitudes of $|D^*|$, we also address how robust the quadrupolar gap is to modifications of the model. We first consider how the $|D^*|$ is affected by the spin value of the impurity $S_{imp}$. For this purpose, we compare the quadrupolar gap at the symmetry point $\varepsilon = -U/2$ obtained for a minimal-spin ($S_{imp} = 1/2$) impurity in Fig. 3(b) of the main article with cases for $S_{imp} = 1, 3/2$ and 2, see Fig. S-9(a). This corresponds to a total high-spin ground state with $S = 3/2$, 2 and 5/2, respectively. We find that with increasing $S_{imp}$ the size of the gap $|D^*|$ also grows, see Fig. S-9(b). Furthermore, our simple model can also be extended to encompass the case of a two-level quantum dot, i.e. by replacing the immobile impurity with a second orbital level for a *mobile* electron. Intra- and inter-level charging energies (both taken equal to $U$) ensure that in the doubly-occupied charge state a single electron resides in each level, and their spins are assumed to be coupled by the same ferromagnetic Heisenberg interaction[101], see Fig. S-9(c). In this model the impurity is thus subject to the most extreme perturbation due to charge fluctuations. Nevertheless, our calculations confirm the general expectation that the quadrupolar gap should appear in any type of system with a well-separated high-spin ground state. In fact, the gap is exactly the same as for the immobile impurity model [compare Figs. S-4(d) and S-9(d)]. The only effect of the charge fluctuations is that the Kondo effect kicks in somewhat earlier, cf. Fig. S-9(d), as evidenced by the significant augmentation of the peaks at the quadrupolar gap in Fig. S-9(c).

We emphasize that our above estimate is still conservative in the following four respects:

1. We used the inflection point of the spectral function to extract the magnitude of the quadrupolar $D^*$ parameter, instead of the peak due to Kondo-enhancement of the inelastic tunneling that overestimates $|D^*|$ (see caption of Fig. S-8).

2. For simplicity we have assumed a half-bandwidth $W$ of $1\,\mathrm{eV}$ as a reference energy scale. Clearly, larger magnitudes of $D^*$ can be achieved when all parameters are rescaled: for a $\lambda$-times larger bandwidth, the value at which $D^*$ saturates is $\lambda$-times larger as well.

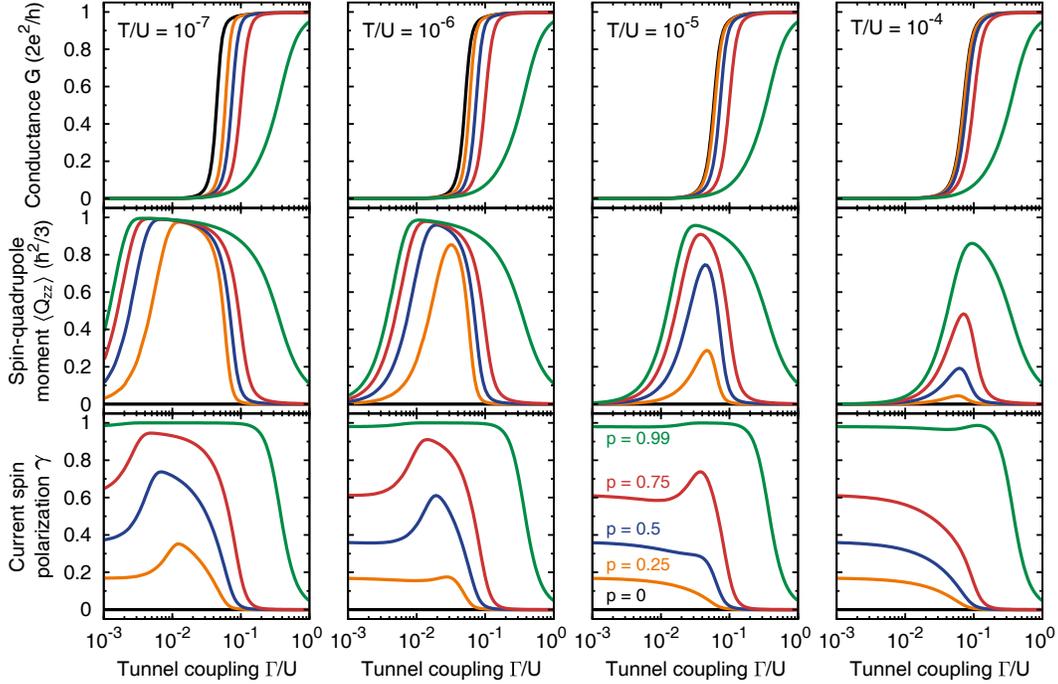

FIG. S-10. **Temperature dependence of spin-anisotropy spin-filtering:** Each figure shows horizontal cuts through Fig. S-5(a),(c)-(d) and how they develop with changing spin-polarization $p$ of the ferromagnets. Parameters are as in Fig. 4 of the main article, except for the temperature (indicated in top figure for each column). (a) Linear conductance $G$, (b) average spin-quadrupole moment $\langle \hat{Q}_{zz} \rangle$ (normalized to its maximal value), (c) linear response spin polarization of transported electrons, $\gamma$, as a function of the tunnel coupling $\Gamma$. Note that the spin-filtering effect vanishes at a temperature scale that is smaller than the scale at which the quadrupolarization $\langle \hat{Q}_{zz} \rangle$ is suppressed, see for example the evolution of the orange curves.

3. We used a minimal value of the impurity spin $S_{\mathrm{imp}} = 1/2$ and therefore the smallest total spin $S = 1$ larger than $1/2$. For larger total spins $S > 1$ the spintronic anistropy gap is larger.

4. Finally, we note that what is estimated above is the *minimal* magnitude of the negative $D$ parameter achieved at the symmetry point, i.e. $D^* = D(\varepsilon = -U/2)$, see Fig. 2 in the main article. One may expect larger magnitudes of $D$ to be accessible when an external magnetic field $B_{\mathrm{ext}}$ is used to compensate the dipolar exchange field $B$. As shown in Ref. [66], this shifts the symmetry point to values of $\varepsilon$ close to the SET resonances where $D(\varepsilon)$ has a larger magnitude while still being negative, see Fig. 2(a). A theoretical analysis of this is, however, beyond the scope of the present study.

## F. Spin-filtering effect

We now discuss the enhancement of the current spin-polarization in Fig. 4 due to the spintronic quadrupolar field and show the temperature-evolution of that figure. First of all, a general indication that the spin-filtering is due to the quadrupolar field stems from the fact that this field is always absent for $S = 1/2$, since in this case $\hat{Q}_{zz} \equiv 0$, i.e. a spin-1/2 system does not 'support' a quadrupolar moment. To be more precise, the crucial difference between spin-1/2 and spin-1 system is the quadrupolar degree of freedom $\hat{Q}_{zz}$, both in the effective Hamiltonian (1), $\mathcal{H}_{\mathrm{eff}} = B\hat{S}_z + D\hat{Q}_{zz}$, and in the reduced density operator describing its state $\rho = \frac{1}{3}\hat{1} + \frac{1}{2}\langle \hat{S}_z \rangle \hat{S}_z + \langle \hat{Q}_{zz} \rangle \hat{Q}_{zz}$, cf. discussion at the end of Sec. II A.

Second, the side panel (ii) of Fig. 4 of the main article illustrates the effective scheme of $S = 1$ energy levels in the superparamagnetic regime. The spintronic anisotropy barrier $|D^*|$ suppresses tunneling-induced spin-flip transitions between the axial states $|S_z = \pm 1\rangle$ and the planar state $|S_z = 0\rangle$, which effectively leads to decoupling of the former two states. In linear response to the bias voltage an asymmetry develops which at low $T \ll |D^*|$ immediately leads to a complete domination of one spin-channel for scattering through the high-spin quantum dot. This high sensitivity to the bias has been discussed in Ref. [48] for intrinsic anisotropy, see in particular Fig. 2 of that reference (compare black dashed and solid lines of the bottom plots). This asymmetry is lifted as soon as the Kondo scale exceeds $|D^*|$ and the both channels start playing a role, reducing ultimately the spin polarization of current to zero. In consequence, it is the possibility of 'taking out' the intermediate planar state $|S_z = 0\rangle$ that prevents the high spin $S \geq 1$ from reversing and thereby allows it to act as a perfect spin-filter. A low-spin $S = 1/2$, in contrast, can always be flipped.

This picture is confirmed by the temperature dependence of the spin-filtering discussed above in Fig. S-5 and Fig. S-6. In Fig. S-10 we show cross-sections of these figures in the same fashion as in Fig. 4 of the main article to illustrate the temperature evolution of that figure. As $T$ is increased, the values of $\Gamma$ required to achieve superparamagnetism $[|D^*(\Gamma)| \geq T$, cf. equation (S-27)] grows, to the point where $T$ and $|D^*(\Gamma)|$ coincide with the Kondo scale and the superparamagnetic state disappears (see Sec. III D), see for example the evolution of the orange curve ($p = 0.25$) in the plot for $\langle \hat{Q}_{zz}\rangle$ in Fig. S-10 when the temperature is increased from the left to the right column. As noted in the caption of Fig. 4 of the article, the spin-filtering effect vanishes at a temperature scale that is smaller than the scale at which the quadrupolarization $\langle \hat{Q}_{zz}\rangle$ is suppressed. There is thus a range of temperatures for which the quadrupolar gap stays intact but the spin filtering effect vanishes. See also the caption of Fig. S-5(d).

## G. On-demand bistability: writing and storing spin

In this section we comment on the electrical control offered by the spintronic dipolar and quadrupolar exchange fields $B$ and $D$, mentioned at the end of the main article. It was experimentally demonstrated by Hauptmann et al.[62] that electrical gating can be used to switch a spin $S = 1/2$ using the dipolar exchange field on a carbon-nanotube quantum dot by tuning from one side of the symmetry point to the other. This type of electric control over the exchange field allows for much faster spin operations than with external magnetic fields . However, by tuning to the symmetry point at $\varepsilon = -U/2$ the written spin could not be safely stored after switching off the exchange magnetic field: at this point the Kondo effect, another consequence of the tunneling coupling to the ferromagnets, strongly perturbs the spin by resonant spin-exchange processes. Our results show that by simply increasing the length of the spin of the quantum-dot to $S = 1$, while keeping the same basic device setup, a well-defined regime around the symmetry point is created where a spintronic magnetic anisotropy barrier protects the spin from environmental perturbations. In Fig. S-11 we schematically depict the stability diagram, as mapped out in Ref. [62], and indicate the quadrupolar-protected regime that is bounded by conditions on the gate-voltage, $|\varepsilon + U/2| \ll \delta\varepsilon = p\Gamma\ln(2W/U)/\pi = \pi|D^*|U/(4p\Gamma)$ , and on the bias voltage $|V_b| \ll |D^*|$.

Keeping a fixed low bias $|V_b| \ll |D^*|$ the spin can thus be rapidly switched by electrically gating from a neutral 'store' point ($\varepsilon = -U/2$) to 'write' a spin 'up' ($\varepsilon < -U/2 - \delta\varepsilon$) or 'down' ($\varepsilon > -U/2 + \delta\varepsilon$), and then returned to the 'store' regime where $|B| \ll |D^*|$ but neither the Kondo effect nor any other perturbations on energy scales less than $|D^*|$ can affect it. We emphasize that both this writing and storing functionality is 'built-in' by simply connecting a quantum dot with $S = 1$ to ferromagnets. This is a striking illustration that cotunneling processes may in fact lead to interesting new functionality (superparamagnetism), instead of just being a limiting factor in device operation[104] (charge noise).

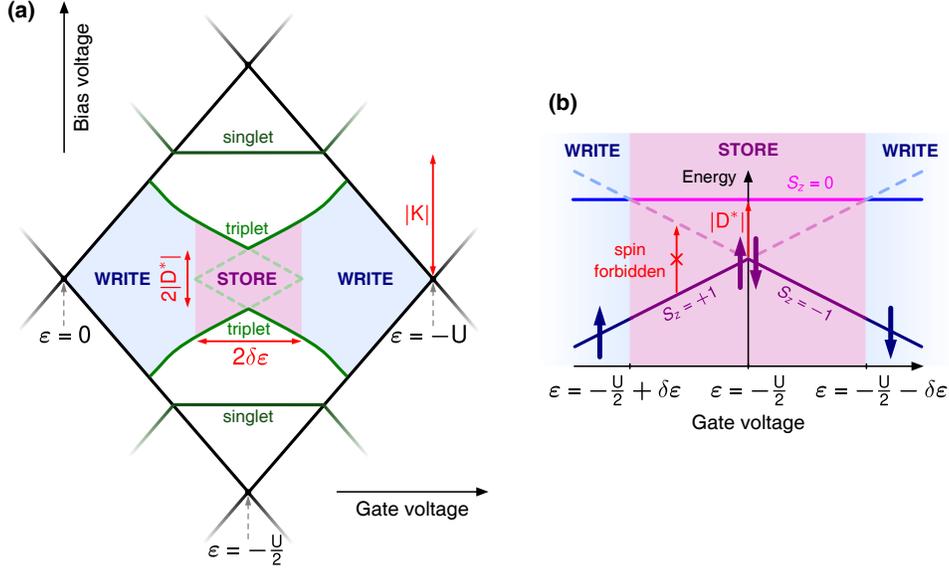

FIG. S-11. **Writing and storing spin:** (a) Schematic stability diagram of the high-spin valve, i.e. a gate- and bias-voltage plane, where for simplicity we take $\varepsilon = -V_{\mathrm{g}}$. Indicated are the Coulomb blockade boundaries (black diamond edges), the excitation energy positions of the high-spin valve [green lines, cf. Fig. 3(c) of the main article], and – for completeness – the singlet inelastic cotunneling and connected SET excitation at $|V_{\mathrm{b}}| = |K|$ (dark green lines). (b) Corresponding energy diagram along the zero-bias line in (a). At the '*store*' point the written axial spin states $|S_z = \pm 1\rangle$ are energetically degenerate, while protected by the quadrupolar gap $|D^*|$ against undesired spin reversal *via* the planar spin state $|S_z = 0\rangle$. Note that for clarity in both schematics the energy scales are exaggerated.

## H. Magnetic switching of spintronic anisotropy, interplay with spin-orbit induced anisotropy and TMR effect

We finally investigate what happens when we switch the ferromagnets' spin-polarization vectors from the *parallel* (P) to the *antiparallel* (AP) configuration. Importantly, the spintronic anisotropy vanishes for the AP configuration as the contributions from the two electrodes cancel each other: the quantum dot then effectively couples to a non-magnetic electrode.[105] One can thus *magnetically* switch off the spintronic anisotropy by reversing the relative alignment of the polarization vectors of the spin-valve device.

This magnetic tunability – in addition to the electric tunability – is a key feature that could possibly be utilized to distinguish the spintronic anisotropy experimentally from the *intrinsic* anisotropy generated by spin-orbit coupling. So far, we focussed our discussion of the quadrupolar gap entirely on nanoscopic systems in which the intrinsic spin anisotropy is negligible. However, the natural question what happens when both *spintronic* ($D$) and intrinsic ($D_{\mathrm{SO}}$) anisotropies are present is a complicated one. Here we merely illustrate the importance of their interplay, we add to equation (S-1) an intrinsic spin anisotropy term $D_{\mathrm{SO}}(s_z + S_{\mathrm{imp}}^z)^2$ with the same signature as the spintronic one, $D_{\mathrm{SO}} \leqslant 0$. Since the intrinsic anisotropy is not affected by the magnetic configuration of the spin-valve, when changing from the P to the AP orientation one expects distinctive features in the conductance that signal the coexistence with the spintronic anisotropy, which is *only* present in the P case. The experimentally relevant quantity to identify changes in transport properties of a system in the two magnetic configurations of electrodes is the tunnel magnetoresistance factor (TMR), defined as

$$\mathrm{TMR} = \frac{G^{\mathrm{P}} - G^{\mathrm{AP}}}{G^{\mathrm{AP}}} \quad \text{with} \quad G^{\mathrm{P/AP}} = G_{\uparrow}^{\mathrm{P/AP}} + G_{\downarrow}^{\mathrm{P/AP}}, \tag{S-32}$$

where $G^{\mathrm{P/AP}}$ ($G_{\sigma}^{\mathrm{P/AP}}$) is the linear conductance (for spin $\sigma = \uparrow, \downarrow$).

In Fig. S-12(a) we present the TMR, calculated using the DM-NRG, as a function of temperature for indicated values of the intrinsic anisotropy $D_{\mathrm{SO}}$. For the case of pure spintronic

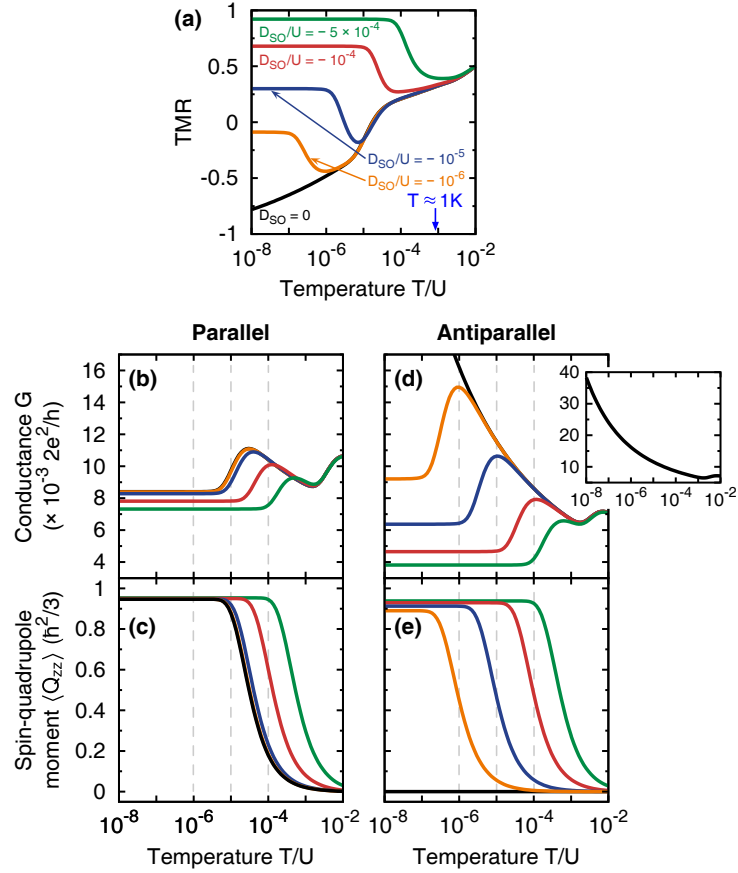

FIG. S-12. **Interplay of spintronic and spin-orbit induced (intrinsic) spin anisotropy:** The tunnel magnetoresistance factor TMR $= (G^{\mathrm{P}} - G^{\mathrm{AP}})/G^{\mathrm{AP}}$ is shown in (a) as a function of temperature $T$ at the symmetry point $\varepsilon = -U/2$ for several values of the intrinsic uniaxial spin anisotropy constant $D_{\mathrm{SO}}$. In the lower panel, the linear conductance $G = G_\uparrow + G_\downarrow$ (b,d), and the quadrupole moment $\langle \hat{Q}_{zz} \rangle$ normalized to its maximal value (c,e) are presented for *parallel* (left column) and *antiparallel* (right column) magnetic configuration of the system. Note that in the pure spintronic limit ($D_{\mathrm{SO}} = 0$) the quadrupole moment is *exactly* zero in the antiparallel configuration [black curve in (e)]. Except for the values of $D_{\mathrm{SO}}$, the temperature $T$ and the polarization $p = 0.75$ all parameters are the same as for Fig. 3(b) in the main article.

anisotropy ($D_{\mathrm{SO}} = 0$, black curves in Fig. S-12), the TMR is a monotonic function of temperature with a smooth step appearing around the temperature where thermal excitations across spintronic quadrupolar gap $|D^*|$ are frozen out. This corresponds to a steep drop of the conductance with decreasing temperature in the P configuration, see Fig. S-12(b), and coincides with the appearance of a finite average quadrupole moment $\langle \hat{Q}_{zz} \rangle$, see Fig. S-12(c). On the contrary, the conductance for the AP orientation does not prominently change at this energy scale, besides showing a smooth increase when the temperature is lowered, see Fig. S-12(d). This reflects the absence of a quadrupolar field in this case, as evidenced by the zero average quadrupole moment[106] $\langle \hat{Q}_{zz} \rangle$ for all temperatures, see Fig. S-12(d).

Clearly, upon introducing intrinsic anisotropy $D_{\mathrm{SO}} \neq 0$ the TMR shows a pronounced qualitative change: upon lowering $T$ the TMR now shows an additional stepwise increase followed by a saturation at a value larger than the TMR value for $D_{\mathrm{SO}} = 0$. Since the intrinsic anisotropy exists in *both* the P and the AP configuration, the conductance reaches a plateau for low temperatures in both cases with, however, different saturation temperatures due to the presence of the spintronic anisotropy only in the P configuration. The corresponding energy scale for the P configuration is set by the combination of intrinsic and spintronic anisotropy, whereas in the AP configuration

the latter is absent. This characteristic two-step behaviour of the TMR, distinctly exemplified, e.g., by the orange curve in Fig. S-12(a), allows the spintronic anisotropy to be detected even in the presence of an intrinsic one of comparable magnitude. We note that for reasons of consistency we have used the theoretically-motivated parameters of the main article. As mentioned there, achievable values of the spintronic anisotropy are conservatively estimated to reach values as large as 0.04 meV and the temperature scales in Fig. S-12 will change accordingly, see Sec. III E.



\end{document}